\documentclass[journal]{IEEEtran}
\ifCLASSINFOpdf
\else
\fi

\usepackage[tbtags]{amsmath}
\usepackage{amsmath,bm}
\usepackage{amssymb}
\usepackage{amsthm}
\usepackage{nomencl}
\usepackage{enumitem}
\usepackage{graphicx}
\usepackage{cite,multirow,gensymb}
\usepackage{subfigure}
\usepackage{mwe}
\usepackage{etoolbox}
\usepackage{booktabs}
\usepackage{caption}
\usepackage{cite}
\usepackage{mathtools}
\usepackage[usenames, dvipsnames]{color}
\usepackage{xcolor}

\usepackage{tabularx,ragged2e,booktabs,caption}
\newcolumntype{C}[1]{>{\Left}m{#1}}

\makeatletter
\patchcmd{\@makecaption}
  {\scshape}
  {}
  {}
  {}
\makeatother

\makenomenclature

\renewcommand\nomgroup[1]{%
    \item[\bfseries
    \ifstrequal{#1}{A}{Sets}{%
    \ifstrequal{#1}{B}{Parameters}{%
    \ifstrequal{#1}{C}{Binary Variables}{%
    \ifstrequal{#1}{D}{Continuous Variables}{}}}}%
]}

\begin{document}
\title{Optimal Transmission Line Switching under Geomagnetic Disturbances}

\author{Mowen Lu$^*$, Harsha Nagarajan$^{\dag}$, Emre Yamangil$^{\dag}$, Russell Bent$^{\ddag}$, Scott Backhaus$^{\ddag}$, Arthur Barnes$^{\ddag}$

\thanks{$*$ Graduate Student, Department of Industrial Engineering, Clemson University, SC, USA}
\thanks{$\dag$ Center for Nonlinear Studies, Los Alamos National Laboratory, NM, USA}
\thanks{$\ddag$ Information Systems and Modeling (A-1 division), Los Alamos National Laboratory, NM, USA}
}


\maketitle

\begin{abstract}
In recent years, there have been increasing concerns about how geomagnetic disturbances (GMDs) impact electrical power systems. Geomagnetically-induced currents (GICs) can saturate transformers, induce hot spot heating and increase reactive power losses. These effects can potentially cause catastrophic damage to transformers and severely impact the ability of a power system to deliver power. To address this problem, we develop a model of GIC impacts to power systems that includes 1) GIC thermal capacity of transformers as a function of normal Alternating Current (AC) and 2) reactive power losses as a function of GIC. We use this model to derive an optimization problem that protects power systems from GIC impacts through line switching, generator redispatch, and load shedding. We employ state-of-the-art convex relaxations of AC power flow equations to lower bound  the objective. We demonstrate the approach on a modified RTS96 system and the UIUC 150-bus system and show that line switching is an effective means to mitigate GIC impacts. We also provide a sensitivity analysis of optimal switching decisions with respect to GMD direction.
\end{abstract}


\IEEEpeerreviewmaketitle



\vspace{-.5cm}

\section{Introduction}
\label{Sec:intro}

\IEEEPARstart{S}{olar} flares and coronal mass ejections drive geomagnetic disturbances (GMD) that lead to changes in the Earth's magnetic field, which then create geo-electric fields. These low-frequency geo-electric fields induce quasi-DC currents, also known as Geomagnetically-Induced Currents (GICs), in grounded sections of power system networks \cite{albertson1973solar,albertso1974effects, albertson1993geomagnetic}. The GIC are superimposed on the usual alternating currents (AC) and bias the AC such that the maximum currents are increased. In many power system components, this bias is not a major concern, however, in transformers, this bias can lead to half-cycle saturation of transformers and the loss of magnetic flux to regions outside of the transformer core. The energy stored in the stray flux increases the reactive power consumption of the transformer, which can affect system voltages. The stray flux also drives eddy currents that can cause excessive transformer heating leading to reduced transformer life or, potentially, immediate damage \cite{GICeffects2012}. 

The potential impacts of GMD to transformers in the bulk electric power system have led the United States government to increase the understanding of and mitigate the impacts of such events\cite{exec2016}, \cite{FEDERALENERGYREGULATORYCOMMISSION2016}. To mitigate the potential risks introduced by GIC to power systems, the electric power industry has actively improved GIC modeling and GIC monitoring \cite{erinmez2002management,cannon2013extreme,qiu2015geomagnetic,overbye2013power, horton2012test, GIC2013flow}. These models have been used to conduct risk analysis \cite{overbye2013power,overbye2012integration} that investigate the sensitivity of transformer reactive power losses due to GIC and concluded that risk and risk mitigation warrants further study. 

One focus in the recent literature has been on mitigating the effect of GIC on transformer reactive power consumption and subsequent drops in system voltages and potential voltage collapse.  One approach to mitigation is the installation of DC-current blocking devices to keep the GIC from entering through transformer neutrals \cite{bolduc2005development}, however, these devices are expensive, with costs for a single unit close to \$500K \cite{liang2015optimal,zhu2015blocking,kovan2015mitigation}. In an attempt to minimize the projected cost of mitigation, optimization-based methods have been developed to guide the siting of these blocking devices. Instead of performing a full power systems analysis that includes the AC, GIC and full AC power flow equations, these papers have primarily focused on minimizing induced reactive losses independent of the normal AC currents. The intuition of these surrogate models is that small amounts of reactive losses imply small voltage impacts and, presumably, a secure power system. 

Beyond voltage effects, the literature on risk mitigation associated with transformer heating is relatively sparse. Existing studies focus on assessing transformer susceptibility to GIC effects \cite{girgis2014process} and formulating the thermal response of transformer cores to different levels of GMDs \cite{marti2013simulation}. However, this approach was strictly a screening study and did not recommend methods for mitigation.

The work discussed above is a very important start, but it leaves a number of open questions, which we address in this manuscript. First, the installation of blocking devices is very expensive and cost may pose a barrier to adoption. Instead, we focus on developing a GIC-aware optimal power flow (OPF) model that uses existing controls such as generator dispatch, load shedding, and line switching to mitigate the risks of GIC impacts. Second, we incorporate the AC physics of power flow into the GIC-aware OPF because these physics play an important role in the impacts associated with GIC. For example, while minimizing reactive losses may imply small voltage problems across the whole system, these models focus on total losses and can miss relatively large voltage problems in a small part of a system. More importantly, models of hotspot thermal heating inherently depend on both GIC and AC.

The setting considered in this manuscript is very challenging. It combines transformer reactive losses, transformer heating, and full AC power flow into an optimization-based operational mitigation setting with line switching. By itself, optimal transmission line switching (OTS) with AC power flow physics is a mathematically challenging problem that includes nonlinearities, nonconvexities and discrete variables. Existing solution methodologies designed for OTS heavily rely on tight convex relaxations and advanced discrete optimization techniques. 
In recent literature, various convex relaxations and disjunctive representations have been developed.
These include second-order-conic (SOC) relaxations \cite{kocuk2017new}, quadratically constrained (QC) relaxations \cite{hijazi2013convex} and Semi-definite programming relaxations \cite{bai2008semidefinite}.
In the context of transmission expansion planning applications, the QC relaxations have been effective \cite{nagarajan2017resilient,nagarajan2016optimal} and we use this model here. Despite these recent advances in optimization methods for OTS, global methods still cannot scale to systems with 500 nodes.

The main contributions of this paper are the formulation and initial algorithmic solution approaches to an operational decision support tool that incorporates:
\begin{enumerate}
    \item A model of transformer heating as a response to AC and GIC-induced DC,
    \item A realistic, coupled model of convex, relaxed AC  power flows with GIC effects and an algorithm to recover good feasible solutions quickly, and 
    \item An optimization problem that protects the system from reactive losses and thermal heating induced by GIC.
\end{enumerate}


\section{GIC modeling and ACOTS formulation}\label{Sec:ACOTS and GIC model}
\nomenclature[A,01]{\color{black}$N^{a}, N^{d}, N^{o} $}{\color{black}set of nodes in the AC and DC circuit, respectively, where $N^o = N^a \cap N^d$}
\nomenclature[A,03]{\color{black}$N^{g} \subseteq N^a$}{set of nodes with exactly one generator}
\nomenclature[A,04]{\color{black}$\mathcal{I} \subseteq N^d$}{\color{black}set of substation neutrals}
\nomenclature[A,07]{\color{black}$\mathcal{E}^a, \mathcal{E}^d, \mathcal{E}$}{\color{black}set of edges in the AC and DC circuit, respectively, where $\mathcal{E} = \mathcal{E}^a \cup \mathcal{E}^d$}
\nomenclature[A,08]{$\mathcal{E}^o \subseteq {\color{black}\mathcal{E}^a}$}{set of transmission lines}
\nomenclature[A,10]{\color{black}$\mathcal{E}^g \subseteq \mathcal{E}^a$}{\color{black}set of edges $e_{ij}$ such that either $i$ or $j$ $\in N^g$}
\nomenclature[A,11]{\color{black}$\mathcal{E}^\tau \subseteq \mathcal{E}^d$}{\color{black}set of transformer edges used to model the high voltage primary windings of GSU transformers and the common windings of autotransformers in the DC circuit.}
\nomenclature[A,12]{\color{black}$\mathcal{E}_i^+\subseteq\mathcal{E}$}{\color{black}set of outgoing edges connected to AC/DC node $i$}
\nomenclature[A,13]{\color{black}$\mathcal{E}_i^-\subseteq\mathcal{E}$}{\color{black}set of incoming edges connected to AC/DC node $i$}
\nomenclature[A,14]{\color{black}$\mathcal{E}_i$}{\color{black}set of all edges connected to AC/DC node $i$, where $\mathcal{E}_i = \mathcal{E}_i^+ \cup \mathcal{E}_i^-$}
\nomenclature[A,15]{\color{black}$\mathcal{E}_i^\tau \subseteq \mathcal{E}^\tau$}{\color{black}set of DC edges used to compute $Q_i^{loss}$ (as described later) for node $i$}

\nomenclature[B,01]{$c_{i}^0,c_{i}^1,c_{i}^2$}{generation cost coefficients of generator $i \in N^g$}
\nomenclature[B,02]{\color{black}$\eta_{e}^0, \eta_{e}^1,\eta_{e}^2$}{coefficients of the thermal limit curve of transformer line ${\color{black}e }
\in \mathcal{E}^{\tau}$}
\nomenclature[B,04]{$\mu$}{cost of load shedding}
\nomenclature[B,05]{\color{black}$a_{m}$}{admittance of the grounding line at bus ${\color{black}m} \in \mathcal{I}$, 0 if bus ${\color{black}m} \not \in \mathcal{I}$} 
\nomenclature[B,06]{\color{black}$a_{e}$}{DC admittance of edge ${\color{black} e }\in {\color{black}\mathcal{E}^d} $}
\nomenclature[B,07]{$J_{e}$}{induced current by GMD on line $ e \in {\color{black}\mathcal{E}^d} $}
\nomenclature[B,08]{$ r_{e}$, $ x_{e} $}{resistance and reactance of line $e \in {\color{black}\mathcal{E}^a} $}
\nomenclature[B,09]{$ g_{e}$, $ b_{e} $}{conductance and susceptance of line $e \in {\color{black}\mathcal{E}^a}  $}
\nomenclature[B,10]{$ g_{i}$, $ b_{i} $}{shunt conductance and susceptance at bus $ i \in {\color{black}N^a} $}
\nomenclature[B,11]{$ d_i^p $, $ d_i^q$}{real and reactive power demand at bus $i \in {\color{black}N^a}$}
\nomenclature[B,12]{$ b_{e}^c $}{line charging susceptance of line $e \in {\color{black}\mathcal{E}^a} $}
\nomenclature[B,13]{$ s_{e} $}{apparent power limit on line $e \in {\color{black}\mathcal{E}^a} $}
\nomenclature[B,14]{$ \overline{\theta}$}{phase angle difference limit}
\nomenclature[B,15]{$ \theta^M$}{Big-M parameter given by $\lvert{\color{black}\mathcal{E}^a}\rvert\overline{\theta}$}
\nomenclature[B,16]{$\overline{I}_{e}^a $}{ AC current flow limit on line $ e \in {\color{black}\mathcal{E}^a} $}
\nomenclature[B,17]{\color{black}$ k_{e} $}{loss factor of transformer line ${\color{black}e} \in \mathcal{E}^{\tau}$}  
\nomenclature[B,18]{$ \underline{V}_i$, $\overline{V}_i$ }{AC voltage limits at bus $i \in {\color{black}N^a}$ }
\nomenclature[B,19]{$ \underline{gp}_{i} $, $\overline{gp}_{i} $}{real power generation limits at generator $i \in G$}
\nomenclature[B,20]{$ \underline{gq}_{i} $, $\overline{gq}_{i} $}{reactive power generation limits at generator $i \in G$}
\nomenclature[B,21]{$\phi$}{the angle of the geo-electric field relative to east}
\nomenclature[B,22]{$\mathcal{V}^d$}{GMD induced voltage source}
\nomenclature[B,23]{$L_N$,$L_E$}{the north and east components of the displacement of each transmission line, respectively}
\nomenclature[B,24]{$E_N$,$E_E$}{strength of the north and east geo-electric field, respectively}
\nomenclature[C,01]{$ z_{e} $}{1 if line {\color{black} $e \in \mathcal{E}^a$} is switched on; 0 otherwise}
\nomenclature[D,01]{$\theta_i$}{phase angle at bus $i \in{\color{black} N^a }$}
\nomenclature[D,04]{$ V_i$}{voltage magnitude at bus $i \in {\color{black}N^a }$}
\nomenclature[D,05]{$V_i^d$}{induced DC voltage magnitude at bus $i \in {\color{black}N^d}$}
\nomenclature[D,06]{$ l_{e}$}{AC magnitude squared on line $e \in {\color{black}\mathcal{E}^a}$}
\nomenclature[D,08]{\color{black}$ I_{e}^d$}{GIC flow on transformer line ${\color{black}e} \in \mathcal{E}^{\tau} $}
\nomenclature[D,09]{$\widetilde{I}_{e}^a$}{AC magnitude on line $e \in {\color{black}\mathcal{E}^a} $}
\nomenclature[D,10]{\color{black}$ \widetilde{I}_{e}^d$}{ the effective GIC on transformer line ${\color{black}e }\in \mathcal{E}^{\tau} $}
\nomenclature[D,11]{$ Q_{i}^{loss}$}{GIC-induced reactive power loss at bus $i \in {\color{black}N^a} $}
\nomenclature[D,12]{$ p_{ij}$, $q_{ij}$}{real and reactive power flow on line $ e_{ij} \in {\color{black}\mathcal{E}^a}$, as measured at node $i$}
\nomenclature[D,13]{$ f_i^p$, $f_i^q$}{real and reactive power generated at bus $i \in {\color{black}N^a}$}
\nomenclature[D,14]{$ l_i^p$, $l_i^q$}{real and reactive power shed at bus $i \in {\color{black}N^a}$}
\printnomenclature[0.6in]

{\color{black} 
Each edge, $e_{ij} \in \mathcal{E}$, is given an arbitrary orientation from bus $i$ to bus $j$. We omit the $ij$ subscript when the orientation is not relevant. For $e \in \mathcal{E}^d$, we use notation $\overrightarrow{e} 
$ to denote the associated AC edge of $e$. This is a one-to-one mapping for transmission lines and a many-to-one mapping for transformers (discussed later).
}

\subsection{GIC Modeling}
\noindent\textbf{$\bm{J_{e}}$ calculation} 
The computation of transformer hot spot heating and GIC-induced reactive power losses depends on the induced current sources ($J_{e}$) on each power line {\color{black} $e \in \mathcal{E}^d$} in the network, which itself depends on the strength and direction of the geo-electric field associated with the GMD. These relationships are modeled in Eq.(\ref{eq:dc_source})
\begin{equation}\label{eq:dc_source}
J_e=a_e\mathcal{V}^d=a_e\oint \vec{E}\cdot d\vec{l},
\end{equation} where, $\vec{E}$ is the geo-electric field at the location of the transmission line, and $d\vec{l}$ is the incremental line segment length, including direction \cite{GIC2013flow}. In practice, the actual geo-electric field varies with time and geographical locations. Using a common assumption that the north and east components of the geo-electric field are constant in the geographical area of the transmission line \cite{horton2012test,zhu2015blocking,GIC2013flow}\footnote{Our model does not depend on this assumption. It only depends on $J_{e}$ as an input parameter.}, $J_{e}$ is calculated as (super- and sub-scripts indicating edges are omitted):
\begin{equation}\label{eq:voltage source}
 \footnotesize
J = a\mathcal{V}^d=a(E_NL_N + E_EL_E) = a|E|(\sin(\phi)L_N + \cos(\phi) L_E),
\end{equation}
where $L_N, L_E, E_N, E_E$ and $\phi$ are as described in the nomenclature (see Appendix I of \cite{GIC2013flow}). {\color{black} Given their short length, generally $J_e = 0$ for transformers, i.e. $e \in \mathcal{E}^\tau$}. 

\noindent\textbf{Transformer modeling}
The two most common transformers in electrical transmission systems subject to GIC are network transformers and generator step-up (GSU) transformers.  Network transformers are generally located relatively far from generators and transform voltage between different sections of the transmission system. In contrast, GSUs connect the output terminals of generators to the transmission network. Many IEEE transmission reliability test networks explicitly model network transformers, but generally do not model GSUs. However, GSUs and the neutral leg ground points they provide are critical when modeling GICs and methods to mitigate the impact of GICs. 

\begin{figure}[htp]
\captionsetup{font=footnotesize}
  \centering
  \subfigure[4-bus system]{
  \includegraphics[scale=0.705]{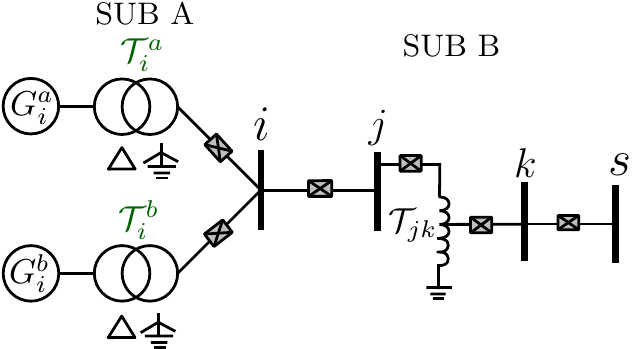}
  \label{fig:4_bus}
  } 
  \subfigure[Equivalent AC network]{
  \includegraphics[scale=0.72]{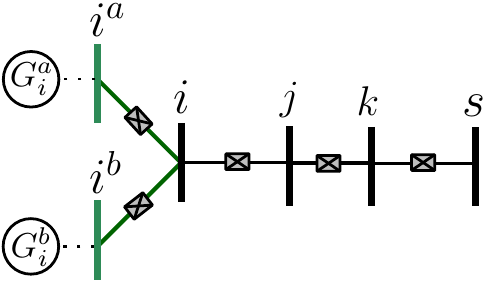}
  \label{fig:4_bus_ac}
  }
  
  \subfigure[Equivalent DC network]{
  \includegraphics[scale=1]{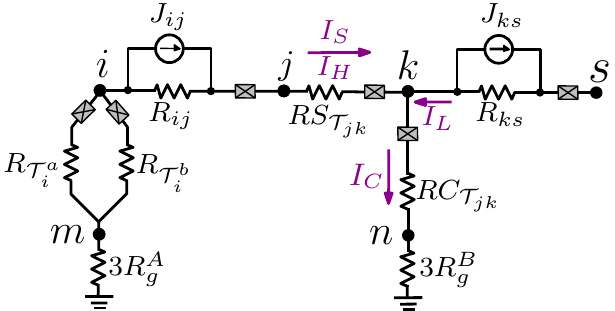}
  \label{fig:4_bus_dc}
  }
\caption{Schematic of the GSU and network transformer modeling used here. \color{black} Picture (c) illustrates the effective GIC calculations. For the autotransformer, $\mathcal{T}_{jk}$, $j$ is the high voltage (HV) bus and $k$ is the low voltage (LV) bus.
$R_g$ denotes the substation grounding resistance. $I_H (I_L)$ and $I_S (I_C)$ are the GIC flows through HV (LV) winding and series (common) winding, respectively.} 
\label{Fig:4_bus_example}
\end{figure}

In this manuscript, we modify the IEEE RTS test network by adding a GSU transformer between each generator and its injection bus (see Fig. \ref{fig:4_bus}). Consistent with common engineering practice, we assume that each GSU is grounded on its high voltage side that connects to the transmission network. We also model the switching of the circuit breaker between the high side of the GSU and the transmission network using a binary variable that allows the GSU to be isolated from the network and the quasi-DC GIC to protect the GSU.  This switching is performed if the generator output is zero. Although the IEEE test networks include network transformers, transformer type and grounding data are typically not provided. In this manuscript, we assume that all network transformers are auto-transformers and each transformer has a single neutral ground on the low voltage side.

Figure \ref{Fig:4_bus_example} includes examples of both GSU and network auto transformer modeling. Figure \ref{fig:4_bus} shows a {\color{black}four-bus} section of the transmission system with a single network transformer ($\mathcal{T}_{jk}$) and two GSU transformers ($\mathcal{T}_i^a,\mathcal{T}_i^b$) independently connecting two generators ($G^a_i,G^b_i$) to the same injection bus $i$. In the simplified AC network of Fig.~\ref{fig:4_bus_ac}, bus $i^a$ and $i^b$ model output terminals of generator $G^a_i$ and $G^b_i$, respectively. Each GSU transformer $\mathcal{T}^a_{i}$ ($\mathcal{T}^a_{j}$) is reduced to a {\color{black} (single)} series impedance $ii^a$ ($ii^b$) with a circuit breaker. The network transformer $\mathcal{T}_{jk}$ is reduced to a {\color{black} (single)} series impedance ($jk$) with a circuit breaker. Under this transformation, the number of buses and lines of {\color{black} the AC network} grow to $|N^o|+|N^g|$ and $|\mathcal{E}^o|+|N^g|$, respectively, where $|N^o|$ and $|\mathcal{E}^o|$ model the original set of buses and edges in the network. Fig.~\ref{fig:4_bus_dc} shows an equivalent single-phase DC circuit of the example system in nodal form. In this figure, {\color{black}$m$ and $n$} model the neutral point of substation $A$ and $B$, respectively. $R_{\mathcal{T}^a_i}$ and $R_{\mathcal{T}^a_i}$ denote the resistance of the primary HV winding of $\mathcal{T}^a_i$ and $\mathcal{T}^b_i$, respectively.\footnote{\color{black} R corresponds to the inverse DC admittance, i.e. $a=\frac{1}{R}$,} $RC_{\mathcal{T}_{jk}}$ and $RS_{\mathcal{T}_{jk}}$ represent the resistance of the common and series windings of $\mathcal{T}_{jk}$, respectively.  For grounded GSU transformers, the effective GIC flows through the primary HV winding. {\color{black} For example, in Fig. \ref{fig:4_bus_dc}, the effective GIC of GSU transformer $\mathcal{T}_i^b$ is  $\widetilde{I}^d_{\mathcal{T}_i^b}$ 
which is the GIC flow from bus $i$ to $m$ on $\mathcal{T}_i^b$}  \cite{zheng2014effects,overbye2012integration}. For an auto-transformer, the effective GIC is derived from the GIC flows through both the series and common windings {\color{black} as shown in Fig. 
\ref{Fig:4_bus_example}(c),
i.e., 

\[\widetilde{I}^d = \left|\frac{\alpha I_H + I_L}{\alpha}\right| = \left|\frac{\alpha_AI_S + I_C}{\alpha_A + 1}\right|\] 

\noindent where $\alpha$ is the turns ratio and $\alpha_A = \alpha_S/\alpha_C = \alpha-1$ (Eq.(14) and Eq.(15)  in \cite{zheng2014effects}). }In this manuscript, we assume the turn ratios of all auto-transformers are one. {\color{black} As a result, $\widetilde{I}^d = |I_H + I_L| = |I_C|$, i.e., the effective GIC is the GIC flow through the common winding.\footnote{The model remains convex for any constant turns ratio by substituting $\left|\frac{\alpha I_H + I_L}{\alpha}\right|$ for $|I_C|$.} Thus, in the four-bus network, the effective GIC of autotransformer $\mathcal{T}_{jk}$ is the GIC flow on line $(k,n)$ of Fig. \ref{fig:4_bus_dc}, i.e., $\widetilde{I}^d_{kn}$. 
} 

\noindent \textbf{GIC-Effects} During GMDs, the quasi-DC GICs may flow through transformers with grounded neutral legs. This quasi-DC current combines with the normal operating AC current creating half-cycle saturation and loss of magnetic flux from the transformer core and leads to several undesirable effects. The two effects that we consider are eddy current-driven transformer heating and excess reactive power consumption from the excess magnetic energy stored in the stray magnetic flux. Both of these effects are challenging to model from first principles, and even if such models existed, they would be too complex to include in the OTS formulation considered here.  Instead, we use a combination of manufacturer test and specification data and simplified models.  

\begin{figure}
\captionsetup{font=footnotesize}
    \centering
    \includegraphics[scale=0.3]{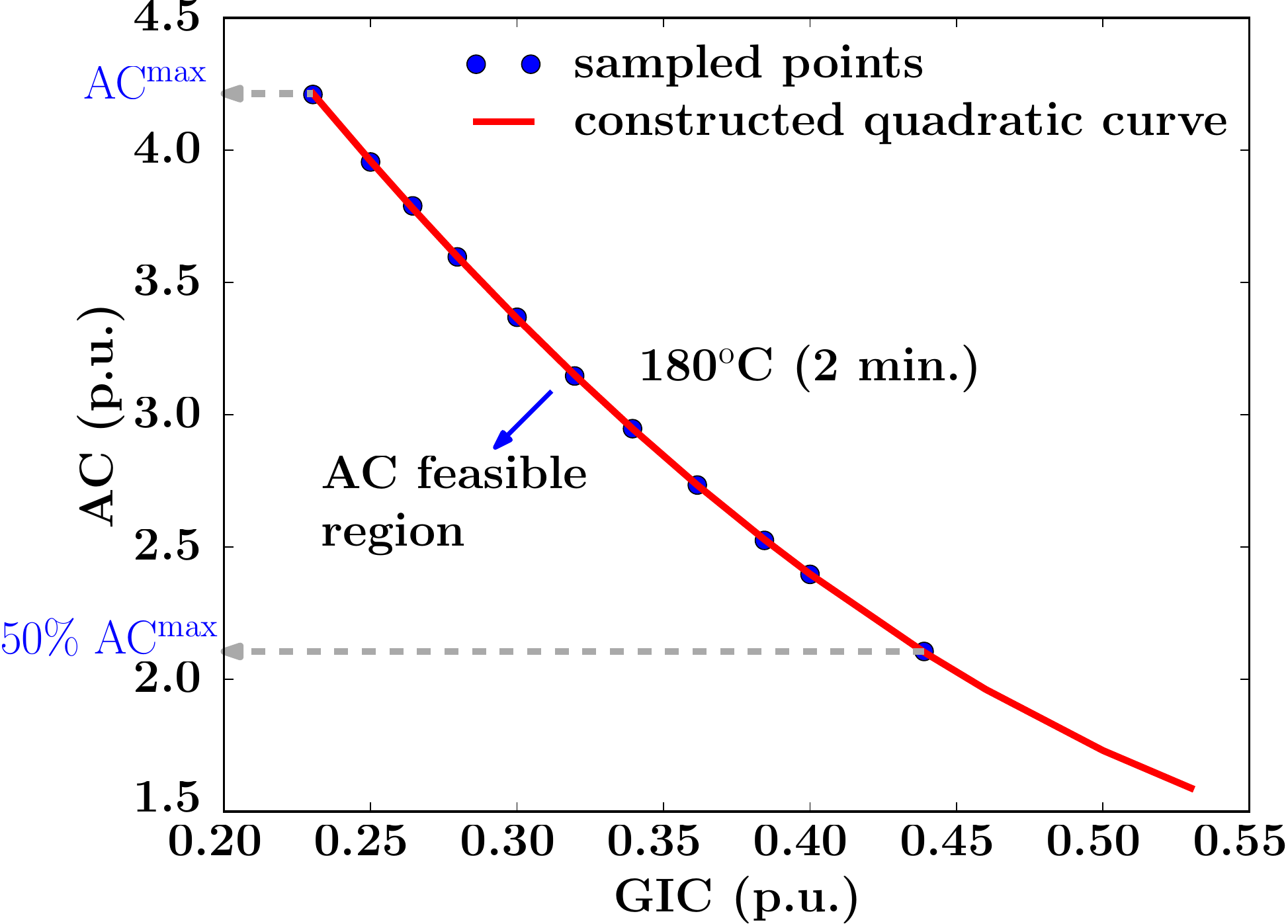}
    \caption{Fitted curve for thermal GIC capability of a transformer. Here, we used 180$\degree$ as the maximum allowed temperature of transformers for short-term (2 minutes) peak GIC pulses and assumed that a transformer cannot be loaded to greater than $100\%$ of its MVA limit. The figure shows the coefficients of the constructed quadratic function (curve), $\eta^0_{e}$, $\eta^1_{e}$ and $\eta^2_{e}$, fitted to the collected 11 (blue) points of the GIC thermal capacity measurements. The feasible region of the transformer load current is the area under the curve and is expressed as constraint (\ref{thermal}).}\label{fig:thermalcurve}
    \vspace{-.7cm}
\end{figure}

For eddy current-driven transformer heating, we use GIC capability curves (e.g. see Fig.~\ref{fig:thermalcurve}) that may be based on either manufacturer acceptance test data or on electromagnetic and thermal modeling of the transformer design. These curves provide an upper bound on a feasible operating range in the space of AC loading and GIC. The upper bound is also a function of the duration of the combined AC and GIC loading (typically given for 30 minute and 2 minute durations). The sampled points (blue) in Fig.~\ref{fig:thermalcurve} are sampled from a transformer manufacturer's 2-minute duration curve \cite{GICcapacity}. Over a reasonable operating range, these points are well represented by the best-fit quadratic (red) curve with the feasible operating region lying below and to the left of the curve. 

Excess reactive power losses due to GIC has been studied in the literature \cite{overbye2012integration,overbye2013power,zhu2015blocking,liang2015optimal}. We adopt the simplified model in \cite{overbye2012integration} which is shown in Eq.(5). These reactive losses create voltage sags that can adversely impact system operation. The previous work has focused on minimizing these losses to improve system safety.  In this manuscript, we explicitly model the AC power equations (voltage magnitudes) so that we can enforce voltage limits directly.

\subsection{ACOTS with GIC constraints}
\label{subsect:NLmodel}

A complete ACOTS model with topology reconfiguration that accounts for GIC-induced transformer thermal heating and transformer reactive power heating is formulated below. 

\begin{subequations}
\label{eq:ACGMD}
\allowdisplaybreaks
\small
\begin{align}
&\label{obj} \min \smashoperator{\sum_{i \in G, e \in \mathcal{E}_i}} c_{i}^2(f^p_i)^2 + c_{i}^1f^p_i + z_{e}(c_{i}^0) + \sum_{i \in N} \mu(l^p_i+l^q_i)\\
& \mathrm{\textbf{AC \ power \ flow \ equations}} \nonumber \\
& \label{pbalance} \smashoperator{\sum_{e_{ij} \in {\color{black}\mathcal{E}_i^+}} } p_{ij} +
\smashoperator{\sum_{e_{ji} \in {\color{black}\mathcal{E}_i^-}} } p_{ij}
=f^p_{i}+l^p_{i}-d^p_{i} - V_i^2g_i \hspace{10pt}\forall i\in {\color{black}N^a}\\
&\label{qbalance}\smashoperator{\sum_{e_{ij}\in {\color{black}\mathcal{E}^+_i}}} q_{ij} +
\smashoperator{\sum_{e_{ji}\in {\color{black}\mathcal{E}^-_i}}} q_{ij}
=f^q_{i}+l^q_{i}-d^q_{i}+V_i^2b_i - Q_i^{loss} \hspace{10pt} \forall i \in {\color{black}N^a}\\
& \label{pij} p_{ij}=z_{e}(g_{e}V_i^2-V_iV_jg_{e} \cos(\theta_i-\theta_j) \nonumber \\ 
& \hspace{25pt} -V_iV_jb_{e} \sin(\theta_i-\theta_j)) \hspace{10pt} \forall e_{ij} \in \mathcal{E}^a \setminus \mathcal{E}^g\\
& \label{qij} q_{ij}=z_{e}(-(b_{e}+\frac{b_{e}^c}{2})V_i^2 + V_iV_jb_{e} \cos(\theta_i-\theta_j) \nonumber \\
& \hspace{25pt} - V_iV_jg_{e} \sin(\theta_i-\theta_j)) \hspace{10pt} \forall e_{ij} \in \mathcal{E}^a \setminus \mathcal{E}^g\\
& \label{pji} p_{ji}=z_{e}(g_{e}V_j^2-V_iV_jg_{e} \cos(\theta_j-\theta_i) \nonumber \\
 & \hspace{25pt} -V_iV_jb_{e} \sin(\theta_j-\theta_i)) \hspace{10pt} \forall e_{ij} \in \mathcal{E}^a \setminus \mathcal{E}^g\\
& \label{qji} q_{ji}=z_{e}(-(b_{e}+\frac{b_{e}^c}{2})V_j^2 + V_iV_jb_{e} \cos(\theta_j-\theta_i) \nonumber \\
 & \hspace{25pt} - V_iV_jg_{e} \sin(\theta_j-\theta_i))  \hspace{10pt} \forall e_{ij} \in \mathcal{E}^a \setminus \mathcal{E}^g\\
& \label{ploss} p_{ij}+p_{ji}=z_{e}r_{e}(l_{e}+b_{e}^c q_{ij}+(\frac{b_{e}^c}{2})^2 V_i^2) \hspace{10pt} \forall e_{ij} \in \mathcal{E}^a \setminus \mathcal{E}^g\\
& \label{qloss} q_{ij}+q_{ji}=z_{e}(x_{e}(l_{e}+b_{e}^c q_{ij}+(\frac{b_{e}^c}{2})^2 V_i^2) \nonumber \\& \hspace{45pt} -\frac{b_{e}^c}{2}(V_i^2+V_j^2)) \hspace{10pt} \forall e_{ij} \in \mathcal{E}^a \setminus \mathcal{E}^g\\
& \label{dummyloss} p_{ij}+p_{ji}= 0, \ q_{ij}+q_{ji}= 0 \hspace{10pt} \forall e_{ij} \in \mathcal{E}^g\\
& \label{sij} p_{ij}^2 + q_{ij}^2 = l_{e} V_i^2 \hspace{10pt} \forall e_{ij} \in {\color{black}\mathcal{E}^a}\\
& \label{lij} l_{e}=(\widetilde{I}_{e}^a)^2 \hspace{10pt} \forall e \in {\color{black}\mathcal{E}^a} \\
& \mathrm{\textbf{Operational \ limit \ constraints}}\nonumber \\ 
& \label{capacity} p_{ij}^2+q_{ij}^2 \leq z_{e}s^2_{e}, \;\;\;p_{ji}^2+q_{ji}^2 \leq z_{e}s^2_{e} \hspace{10pt} \forall e_{ij}  \in {\color{black}\mathcal{E}^a}\\
& \label{Ia} 0 \leq \widetilde{I}_{e}^a \leq z_{e}\overline{I}_{e}^a \hspace{10pt} \forall e \in {\color{black}\mathcal{E}^a} \\
& \label{vi}\underline{V}_i \leq V_i \leq \overline{V}_i \hspace{10pt} \forall i \in {\color{black}N^a}\\
& \label{thetaij_ub} |\theta_i-\theta_j| \leq z_{e}\overline{\theta} + (1-z_{e})\theta^M \hspace{10pt} \forall e_{ij} \in {\color{black}\mathcal{E}^a \setminus \mathcal{E}^g} \\
& \label{gp} z_{e}\underline{gp}_i \leq f_i^p \leq z_{e}\overline{gp}_i \hspace{10pt} \forall i \in G,\hspace{3pt} e \in \mathcal{E}_i\\
& \label{gq} z_{e}\underline{gq}_i \leq f_i^q \leq z_{e}\overline{gq}_i \hspace{10pt} \forall i \in G, \hspace{3pt} e \in \mathcal{E}_i \\
& \mathrm{\textbf{GIC \ effects \ on \ transformers}} \nonumber \\
&\nonumber\smashoperator{ \sum_{e \in \mathcal{E}^+_m} }z_{\overrightarrow{e}}J_{e}
-
\color{black} \smashoperator{\sum_{e \in \mathcal{E}^-_m}} z_{\overrightarrow{e}}J_{e}
= -a_{m}V_m^d \\&\label{GIC}-\smashoperator{\sum_{e_{mn} \in \mathcal{E}_m^+}} z_{e}a_{e}(V_m^d - V_n^d)
+ \ \smashoperator{\sum_{e_{nm} \in \mathcal{E}_m^-}} z_{e}a_{e}(V_n^d - V_m^d)
\;\;\forall m \in N^d \hspace{-0pt}\\
& \color{black}\label{Id} I_{e}^d=z_{\overrightarrow{e}}a_{e}(V_m^d-V_n^d) \hspace{10pt} \forall e_{mn} \in \mathcal{E}^{\tau} \\
& \color{black}\label{Idmag} \widetilde{I}_{e}^d \geq I_{e}^d, \hspace{10pt} \widetilde{I}_{e}^d \geq - I_{e}^d \hspace{10pt} \hspace{10pt} \forall e \in \mathcal{E}^{\tau} \\
& \label{Idub} 0 \leq {\color{black}\widetilde{I}_{e}^d }\leq \max_{\forall \hat{e} \in {\color{black}\mathcal{E}^a}} 2\overline{I}_{\hat{e}}^a \hspace{10pt} {\color{black}\forall e \in \mathcal{E}^{\tau} }\\
& \label{thermal} \widetilde{I}_{\overrightarrow{e}}^a \leq {\color{black}\eta^0_{e}+\eta^1_{e}\widetilde{I}_{e}^d + \eta^2_{e}(\widetilde{I}_{e}^d)^2 \hspace{10pt} \forall e \in  \mathcal{E}^\tau }\\
& \color{black}\label{Qloss} Q_i^{loss}=\smashoperator{ \sum_{e \in \mathcal{E}_i^{\tau}}}k_{e}V_i\widetilde{I}^d_{e} \hspace{10pt}\forall i \in N^a \\
& \label{binary} z_{e} \in \{0, 1 \} \hspace{10pt} \forall e \in {\color{black}\mathcal{E}^a}
\end{align}
\end{subequations}

\vspace{-.2cm}
The objective function (\ref{obj}) minimizes total generator dispatch costs and load shedding costs.{ \color{black} Constraints (\ref{pbalance}) -- (\ref{gq}) describe system constraints for the buses and branches in the AC circuit.} Constraints (\ref{pbalance}) and (\ref{qbalance}) represent the nodal real and reactive power balance, including the increased reactive power losses (demand) due to GICs. Constraints (\ref{pij}) through (\ref{qji}) model the AC power flow on each transmission line with on-off variables {\color{black}$z_{e}$}. The flow on any line is forced to zero if the line is switched off. Constraints (\ref{ploss}) through (\ref{dummyloss}) model power loss equations associated with AC power flow. In constraint \eqref{dummyloss},  
\textit{fictitious} lines between output terminals of generators and their injection buses are modeled as transportation edges (i.e., $|p_{ij}|=|-p_{ji}|= f^p_i, |q_{ij}|=|-q_{ji}|= f^q_i \hspace{6pt} \forall e_{ij} \in \mathcal{E}^g $). Nonconvex constraint (\ref{sij}) evaluates current magnitude $l_{ij}$, an auxiliary variable introduced to bound the squared AC current flow magnitude in constraint (\ref{lij}). Constraints (\ref{capacity}) through (\ref{gq}) describe the operational limits of the grid; constraint (\ref{capacity}) models operational thermal limits of lines in both directions. Constraint (\ref{vi}) limits the voltage magnitude at buses. Constraint (\ref{thetaij_ub}) applies appropriate bounds on phase angle difference between two buses when the line exists. 
Constraints (\ref{gp}) and (\ref{gq}) model the availability and capacity of power generation. A generator is offline if its line is switched off. 

{\color{black}The DC circuit and the effects associated with the GMD are formulated in constraints (\ref{GIC})-(\ref{thermal}). Recall that we link an edge, $e \in \mathcal{E}^d$ in the DC circuit to an edge in the AC circuit with $\overrightarrow{e}$.
Also recall that the HV primary winding of GSU transformer $e_{ij} \in \mathcal{E}^a$ is modeled by introducing a node and edge in the DC circuit (node $m$ in Fig. \ref{Fig:4_bus_example}(c)). 
Similarly, the common winding of autotransformer $e_{ij} \in \mathcal{E}^a$ is modeled in the DC circuit by introducing additional nodes and edges (see Fig. \ref{Fig:4_bus_example}(c)). By using these notations,} constraints (\ref{GIC}) and (\ref{Id}) calculate the GIC {\color{black}flow on each DC} line by applying Kirchhoff's current law. The GIC on a line is determined by the induced current source and the quasi-dc voltage difference between two buses \cite{GIC2013flow}. GIC flow is forced to 0 by $z_{e}$ when $e$ is switched off. Since the value of {\color{black}$I_{e}^d$} can be negative, decision variables {\color{black}$\widetilde{I}_{e}^d$} are introduced to model the magnitude (absolute value) of {\color{black}GIC flows (i.e., $\widetilde{I}_{e}^d = |I_{e}^d|$)}. Instead of introducing additional discrete variables, constraint (\ref{Idmag}) is used to model and relax the magnitude of {\color{black}$I_{e}^d$}. Constraint (\ref{Idub}) denotes the maximum allowed value of GIC flowing through transformers. We assume this limit is twice the upper bound of AC flows in the network. Constraint (\ref{thermal}) guarantees that the hot spot temperature of transformers due to the combination of AC and GICs is below the thermal limits for peak GIC. 
Constraint (\ref{Qloss}) computes the reactive power load due to GIC transformer saturation \cite{albertson1973solar,overbye2013power, zheng2014effects, zhu2015blocking} {\color{black} by using the effective GIC on the primary winding in GSU transformers and the common winding in autotransformers ($\mathcal{E}^\tau_i$)}. The couplings between AC power flows and GIC
occur in constraints (\ref{qbalance}), (\ref{thermal}), and (\ref{Qloss}).

\vspace{-0.3cm}
\subsection{Convex Relaxations}

\label{subsec:ConvexRelaxation}

The ACOTS with GIC constraints is a mixed-integer, non-convex optimization problem that is generally computationally very difficult to solve. 
We adopt the convex relaxations developed by \cite{hijazi2013convex} and later show that the results obtained with the relaxation is (empirically) tight. We now discuss the key features of the relaxations extended to the problem with GIC.

\noindent \textbf{Handling bilinear terms}
Given any two variables $x_i$, $x_j \in \mathbb{R}$, the McCormick relaxation is used to linearize the bilinear product $x_ix_j$ by introducing a new variable $\widehat{x_{ij}} \in {\langle x_i, x_j \rangle}^{MC}$. The feasible region of $\widehat{x_{ij}}$ is defined by inequalities (\ref{eq:SMC}). Note that the MC relaxation is exact if one variable is binary. 

\vspace{-0.5cm}

\begin{subequations} \label{eq:SMC}
\small
\allowdisplaybreaks
\begin{align}
& \label{McCormick}\widehat{x_{ij}} \geq \underline{x}_ix_j+\underline{x}_jx_i -\underline{x}_i \hspace{2pt} \underline{x}_j \\ 
&\widehat{x_{ij}} \geq \overline{x}_ix_j+\overline{x}_jx_i - \overline{x}_i \hspace{2pt} \overline{x}_j \\ 
&\widehat{x_{ij}} \leq \underline{x}_ix_j+\overline{x}_jx_i-\underline{x}_i \hspace{2pt} \overline{x}_j \\ 
&\widehat{x_{ij}} \leq \overline{x}_ix_j+\underline{x}_jx_i-\overline{x}_i \hspace{2pt} \underline{x}_j \\
& \underline{x}_i \leq x_i \leq \overline{x}_i, \ \underline{x}_j \leq x_j \leq \overline{x}_j
\end{align}
\end{subequations}%

\noindent \textbf{Quadratic terms}
Given a variable $x_i \in \mathbb{R}$, a second-order conic relaxation can be applied to convexify the quadratic term $x_i^2$ by introducing a new variable $\widehat{x_{i}} \in {\langle x_i\rangle}^{MC-q}$, as defined in equation (\ref{eq:MC-q}). 
\vspace{-.25cm}
\begin{subequations} 
\label{eq:MC-q}
\small
\allowdisplaybreaks
\begin{align}
& \label{McC-q}\widehat{x_{i}} \geq x_i^2\\ 
&\widehat{x_{i}} \leq (\overline{x}_i+\underline{x}_i)x_i - \overline{x}_i\underline{x}_i\\ 
& \underline{x}_i \leq x_i \leq \overline{x}_i
\end{align}
\end{subequations}%

\noindent \textbf{On/off trigonometric terms}
In constraints \eqref{pij}, \eqref{qij}, \eqref{pji} and \eqref{qji}, if the line $e_{ij}$ is switched off, \cite{hijazi2013convex} suggests the following procedure to deactivate the associated trigonometric terms:
Given the phase angle difference variable $\theta_{ij} = \theta_i - \theta_j$ and on/off variable $z_{e} \in \{0, 1\}$, a disjunctive quadratic relaxation is used to convexify the nonlinear function $z_{e} \cos(\theta_{ij})$ by introducing a new variable $\widehat{cs}_{ij} \in {\langle z_{e} cos(\theta_{ij})\rangle}^{R}$, as formulated in (\ref{eq:cos}).

\begin{subequations} 
\label{eq:cos}
\small
\allowdisplaybreaks
\begin{align}
&\widehat{cs}_{ij} \leq z_{e} - \frac{1- \cos(\overline{\theta})}{(\overline{\theta})^2}(\theta_{ij}^2+(z_{e}-1)(\theta^{u})^2) \\
&z_{e}\cos(\overline{\theta}) \leq \widehat{cs}_{ij} \leq z_{e}
\end{align}
\end{subequations}%

Similarly, for $z_{e}sin(\theta_{ij})$, a disjunctive polyhedral relaxation is applied by introducing a new variable $\widehat{s}_{ij} \in {\langle z_{e} sin(\theta_{ij})\rangle}^{R}$, as described in equation (\ref{eq:sin}).
\begin{subequations} 
\label{eq:sin}
\small
\allowdisplaybreaks
\begin{align}
& \widehat{s}_{ij} \leq \cos(\bar{\theta}/2)\theta_{ij}+z_{e}(\sin(\bar{\theta}/2)-\bar{\theta}/2\cos(\bar{\theta}/2)) \\ & \nonumber \hspace{20pt}+ (1-z_{e})(\cos(\bar{\theta}/2)\theta^{M} +1)\\
&\widehat{s}_{ij} \geq \cos(\bar{\theta}/2)\theta_{ij}-z_{e}(\sin(\bar{\theta}/2)-\bar{\theta}/2\cos(\bar{\theta}/2)) \\ & \nonumber \hspace{20pt}- (1-z_{e})(\cos(\bar{\theta}/2)\theta^{M} +1)\\
&z_{e}\sin(-\overline{\theta}) \leq \widehat{s}_{ij} \leq z_{e}\sin(\overline{\theta})
\end{align}
\end{subequations}%

Based on the above relaxations, we replace
the non-convex constraints in \eqref{pij}, \eqref{qij}, \eqref{pji} and \eqref{qji} with equations (\ref{Quad}):

\begin{subequations}
\label{Quad}
\small
\begin{align}
&  p_{ij}=g_{e} \widehat{zv}_{ij}-g_{e} \widehat{wc}_{ij} - b_{e} \widehat{ws}_{ij} \hspace{10pt} \forall e_{ij} \in \mathcal{E}^a \setminus \mathcal{E}^g\\
& q_{ij}=-(b_{e}+\frac{b_{e}^c}{2}) \widehat{zv}_{ij} + b_{e}\widehat{wc}_{ij} - g_{ij} \widehat{ws}_{ij} \hspace{4pt} \forall e_{ij} \in \mathcal{E}^a \setminus \mathcal{E}^g
\end{align}
\end{subequations}
where, the new variables $\widehat{zv_{ij}}$, $\widehat{wc}_{ij}$ and $\widehat{ws}_{ij}$, admit feasible regions as given in equations (\ref{PFRelax}):
\begin{subequations}
\label{PFRelax}
\small
\begin{align}
& \widehat{cs}_{ij} \in {\langle z_{e}cos(\theta_{ij})\rangle}^{R}, \ \ \widehat{s}_{ij} \in {\langle z_{e}sin(\theta_{ij})\rangle}^{R}\\
& \widehat{v_i} \in {\langle V_i\rangle}^{MC-q}, \ \ \widehat{zv}_{ij} \in {\langle z_{ij}, \widehat{v}_i \rangle}^{MC}, \\ 
&\widehat{w}_{ij}\in {\langle V_i, V_j \rangle}^{MC},\\
& \widehat{wc}_{ij} \in {\langle \widehat{w}_{ij}, \widehat{cs}_{ij}\rangle}^{MC}, \ \ \widehat{ws}_{ij} \in {\langle \widehat{w}_{ij}, \widehat{s}_{ij} \rangle}^{MC}
\end{align}
\end{subequations}

\noindent \textbf{Other nonconvex constraints}
Further, non-convex constraints \eqref{sij} and \eqref{lij} are relaxed to a convex, rotated second-order conic constraint by using the introduced lifted variable $\widehat{v_i}$ (for \eqref{sij}) as follows: 
\begin{subequations}
\begin{align}
    &p_{ij}^2 + q_{ij}^2 \leq l_{e}\widehat{v_i} \ \ \forall e_{ij} \in {\color{black}\mathcal{E}^a}, \\
    & l_{e}\geq (\widetilde{I}_{e}^a)^2 , \  l_{e}\leq (\overline{I}_{e}^a)\widetilde{I}_{e}^a\ \ \forall e \in {\color{black}\mathcal{E}^a}
\end{align}
\end{subequations}

Convex relaxations of the reformulated thermal heating limit constraint (\ref{thermal}) and excess reactive power losses equation (\ref{Qloss}) are stated here in equations
(\ref{eq:thermal&qloss}):
\begin{subequations}
\label{eq:thermal&qloss}
\small
\begin{align}
& \widetilde{I}_{\overrightarrow{e}}^a \leq {\color{black}\eta^0_{e}+\eta^1_{e}\widetilde{I}_{e}^d + \eta^2_{e}\widehat{I}_{e}^d \hspace{10pt} \forall e \in  \mathcal{E}^\tau }\\
& Q_i^{loss}={\color{black}\smashoperator{ \sum_{ e \in \mathcal{E}_i^{\tau}}}k_{e}\widehat{VI}^d_{e} \hspace{10pt}\forall i \in N^a }\\
&\color{black} \widehat{I}_{e}^d \in {\langle \widetilde{I}_{e}^d \rangle}^{MC-q} \hspace{10pt} \forall e \in  \mathcal{E}^\tau \\ &\color{black}\widehat{VI}_{e}^d \in {\langle {V}_{i}, \widetilde{I}_{e}^d \rangle}^{MC} \hspace{10pt} 
\forall i \in \mathcal{E}^a, \ \ \forall e \in  \mathcal{E}^\tau_i
\end{align}
\end{subequations}%

{\color{black}
Convex relaxations of the GIC injection constraints (\ref{GIC}) and (\ref{Id}) are described in equation (\ref{eq:gic_mc}).
\begin{subequations}
\label{eq:gic_mc}
\small
\begin{align}
&\nonumber \smashoperator{\sum_{e \in \mathcal{E}^+_m}}z_{\overrightarrow{e}}J_{{e}} 
-\smashoperator{\sum_{e \in \mathcal{E}^-_m}}z_{\overrightarrow{e}}J_{{e}}
= -a_{m}V_m^d \\ 
&-\smashoperator{\sum_{e \in \mathcal{E}^+_m}}a_{e}\widehat{zv}_{e}^d
+\smashoperator{\sum_{e \in \mathcal{E}^-_m}}a_{e}\widehat{zv}_{e}^d
\hspace{8pt} \forall m \in N^d \\ 
& I_{e}^d=a_{e}\widehat{zv}_{e}^d \hspace{10pt} \forall e \in \mathcal{E}^{\tau} \\
& \widehat{zv}_{e}^d \in {\langle {z}_{\overrightarrow{e}}, (V^d_m - V^d_n) \rangle}^{MC} \hspace{10pt}\forall e_{mn} \in \mathcal{E}^d
\end{align}
\end{subequations}%
}


\vspace{-.5cm}

\section{Case Study}
\label{Sec:case study}

In this section, we analyze the performance and sensitivity of a power system when exposed to varying strengths of geo-electric fields induced by GMDs. 
We use a modified version of single area IEEE RTS-96 system \cite{wong1999ieee}.
Its size is comparable to previous work \cite{zhu2015blocking} that considered minimization of the quasi-static GICs and not a full AC-OPF with topology control.
The derived and modified parameters of IEEE RTS-96 are presented in Table \ref{tb:para_1}--\ref{tb:para_2}. We arbitrarily placed the system in western Pennsylvania to give the model a geographic orientation. We assume the cost of shedding load is twice the cost of the most expensive generator. We performed all computations using the high performance computing resources at Los Alamos National Laboratory with Intel(R) Xeon(R) CPU E5-2660 v3 @2.60GHz processor and 120 GB memory installed. All cases were solved using CPLEX 12.7.0 (default options). Knitro 10.2.1 (default options) was used as the local solver. JuMP was used as an algebraic modeling language \cite{dunning2015jump}.

For reference, the peak geo-electric field during the HydroQuebec event of 1989 was 2 V/km ($\approx$3.2 V/mile) \cite{walling1991characteristics, boteler1994geomagnetically}. References \cite{pulkkinen2012generation} and \cite{backhaus} suggests that 100-year GMDs could cause 5 V/km ($\approx$9 V/mile) and 13 V/km  ($\approx$21 V/mile), respectively, at some high-latitude locations. In our case studies, we consider middle ground, but still extreme, geo-electric fields of 12 V/mile and 14 V/mile. We also study the directionality of the event by considering field directions between $0\degree$ and $360\degree$ spaced by $10\degree$.

To analyze the benefits for GIC mitigation of generator dispatch and load shedding and the combined effects of those two controls plus topology control, we studied three cases. To describe these cases, we define $\mathbf{z}^*_x$ and $c^*_x$ to be the optimal topology (line on/off) decisions and objective (minimum total costs), respectively, for case $x$. The models are defined below as ($M_x$). 
The solutions of $\mathbf{z}_x^*$ and $c_x^*$ are obtained from the convex relaxations of $M_x$ described in Section \ref{subsec:ConvexRelaxation}. The cases we consider are:

\begin{enumerate}
  \item C1: The ACOTS model neglecting GIC effects ($c^*_{o}$, $\mathbf{z}^*_{o}$): \\
    $M_o := $ Min\{(\ref{obj}): (\ref{pbalance})-(\ref{gq}); (\ref{binary}); $\mathbf{Q}^{loss} = \mathbf{0}$\}
  \label{case1}
  \item C2: The ACOTS with GIC effects ( $c^*_{gmd}$, $\mathbf{z}^*_{gmd}$):\\
  $M_{gmd} := $ Min\{(\ref{obj}): (\ref{pbalance})-(\ref{binary})\}
  \label{case2}
  \item C3: The ACOPF (fixed $\mathbf{z}=\mathbf{z}^*_{o}$) with GIC effects ($c^*_{f}$, $\mathbf{z}^*_{o}$):\\
  $M_{f} := $ Min\{(\ref{obj}): (\ref{pbalance})-(\ref{binary}); $\mathbf{z}=\mathbf{z}^*_{o}$\}
  \label{case3}
  \item C4: The ACOPF (fixed $\mathbf{z}=1$) with GIC effects ($c^*_{\mathbf{1}}$, $\mathbf{1}$):\\
  $M_{\mathbf{1}} := $ Min\{(\ref{obj}): (\ref{pbalance})-(\ref{binary}); $\mathbf{z}=1$\}
  \label{case4}
\end{enumerate}%

Case C1 defines the topology $\mathbf{z}^*_o$ and evaluates the objective $c^*_o$ that results from neglecting the effects of GICs. Case C3 evaluates the new cost $c^*_f$ that results from mitigating GIC with generation dispatch on the topology of C1. Case C4 is similar to case C3, but all lines are closed. Case C2 considers the effects of both generation dispatch and topology control. {\color{black} All results in this section are based on the convex relaxation (except for Section \ref{sec:recover} which evaluates the quality of the relaxation by recovering feasible solutions to the original non-convex formulation).}

\begin{table}
  \captionsetup{font=footnotesize}
  \centering
  \scriptsize
  \caption{Transformer data. In this manuscript, all network transformers are auto-transformers. The transformer winding resistance and $k$ are estimated based on the test cases provided in \cite{GIC2013flow, horton2012test}. }
  \setlength{\tabcolsep}{0.85em}
  \allowbreak
  \begin{tabular}{lccccccc}
  \toprule
  &&Resistance&&Resistance&&&\\
  Name&Type&W1&Bus&W2&Bus&Line&k\\
  &&(Ohm)&No.&(Ohm)&No.&No.&(p.u.)\\
  \midrule
A 1&Auto&0.12&3&0.18&24&7&1.8\\
A 2&Auto&0.12&9&0.18&11&14&1.8\\
A 3&Auto&0.12&9&0.18&12&15&1.8\\
A 4&Auto&0.12&10&0.18&11&16&1.8\\
A 5&Auto&0.12&10&0.18&12&17&1.8\\
G 1&GSU&0.3&1& N/A &25&44&1.8\\
G 2&GSU&0.3&1& N/A &26&45&1.8\\
G 3&GSU&0.3&1& N/A &27&46&1.8\\
G 4&GSU&0.3&1& N/A &28&47&1.8\\
G 5&GSU&0.3&2& N/A &29&48&1.8\\
G 6&GSU&0.3&2& N/A &30&49&1.8\\
G 7&GSU&0.3&2& N/A &31&50&1.8\\
G 8&GSU&0.3&2& N/A &32&51&1.8\\
G 9&GSU&0.3&7& N/A &33&52&1.8\\
G 10&GSU&0.3&7& N/A &34&53&1.8\\
G 11&GSU&0.3&7& N/A &35&54&1.8\\
G 12&GSU&0.3&13& N/A &36&55&1.8\\
G 13&GSU&0.3&13& N/A &37&56&1.8\\
G 14&GSU&0.3&13& N/A &38&57&1.8\\
G 15&GSU&0.3&14& N/A &39&58&1.8\\
G 16&GSU&0.3&15& N/A &40&59&1.8\\
G 17&GSU&0.3&15& N/A &41&60&1.8\\
G 18&GSU&0.3&15& N/A &42&61&1.8\\
G 19&GSU&0.3&15& N/A &43&62&1.8\\
G 20&GSU&0.3&15& N/A &44&63&1.8\\
G 21&GSU&0.3&15& N/A &45&64&1.8\\
G 22&GSU&0.3&16& N/A &46&65&1.8\\
G 23&GSU&0.3&18& N/A &47&66&1.8\\
G 24&GSU&0.3&21& N/A &48&67&1.8\\
G 25&GSU&0.3&22& N/A &49&68&1.8\\
G 26&GSU&0.3&22& N/A &50&69&1.8\\
G 27&GSU&0.3&22& N/A &51&70&1.8\\
G 28&GSU&0.3&22& N/A &52&71&1.8\\
G 29&GSU&0.3&22& N/A &53&72&1.8\\
G 30&GSU&0.3&22& N/A &54&73&1.8\\
G 31&GSU&0.3&23& N/A &55&74&1.8\\
G 32&GSU&0.3&23& N/A &56&75&1.8\\
G 33&GSU&0.3&23& N/A &57&76&1.8\\
  \bottomrule
  \end{tabular}
\label{tb:para_1}
\end{table}

\begin{table}
  \captionsetup{font=footnotesize}
  \caption{Power system model parameters. The nominal line length parameters of a single area of RTS-96 \cite{wong1999ieee} are used to perform an approximate geospatial layout of the power system nodes. (b) The substation grounding resistance $GR$ is estimated from typical values of grounding resistance of substations provided in \cite{morstad2012grounding}. (c) The original line parameters $r_{e}^o$ and $x_{e}^o$ are scaled by the ratio $kl_{e}$ of the new to original line lengths.}
  \centering
  \scriptsize
  \subtable[Transmission line data]{
  \setlength{\tabcolsep}{0.6em}
  \begin{tabular}{lcccc}
  \toprule
    Line & From & To & Length \\
    & Bus & Bus & (miles) \\
    \midrule
    1 & 1 & 2 & 3.98 \\
    2 & 1 & 3 & 53.15 \\
    3 & 1 & 5 & 22.78 \\
    4 & 2 & 4 & 33.16 \\
    5 & 2 & 6 & 44.49 \\
    6 & 3 & 9 & 33.56 \\
    7 & 3 & 24 & 0.00 \\
    8 & 4 & 9 & 26.89 \\
    9 & 5 & 10 & 23.38 \\
    10 & 6 & 10 & 19.96 \\
    11 & 7 & 8 & 16.04 \\
    12 & 8 & 9 & 43.51 \\
    13 & 8 & 10 & 43.51 \\
    14 & 9 & 11 & 0.00 \\
    15 & 9 & 12 & 0.00 \\
    16 & 10 & 11 & 0.00 \\
    17 & 10 & 12 & 0.00 \\
    18 & 11 & 13 & 35.95 \\
    19 & 11 & 14 & 33.98 \\
    20 & 12 & 13 & 35.95 \\
    21 & 12 & 23 & 70.48 \\
    22 & 13 & 23 & 57.39 \\
    23 & 14 & 16 & 27.36 \\
    24 & 15 & 16 & 12.18 \\
    25 & 15 & 21 & 35.44 \\
    26 & 15 & 21 & 35.44 \\
    27 & 15 & 24 & 38.43 \\
    28 & 16 & 17 & 18.77 \\
    29 & 16 & 19 & 18.57 \\
    30 & 17 & 18 & 10.75 \\
    31 & 17 & 22 & 72.84 \\
    32 & 18 & 21 & 17.96 \\
    33 & 18 & 21 & 17.96 \\
    34 & 19 & 20 & 29.97 \\
    35 & 19 & 20 & 29.97 \\
    36 & 20 & 23 & 15.59 \\
    37 & 20 & 23 & 15.59 \\
    38 & 21 & 22 & 51.83 \\
  \bottomrule
  \end{tabular}
  }
  \subtable[Substation data]{
  \setlength{\tabcolsep}{0.4em}
  \begin{tabular}{lcccc}
  \toprule
    &&&GR\\
    Name&Latitude&Longitude&(Ohm)\\
  \midrule
    SUB 1&40.44&-78.80&0.1 \\
    SUB 2&40.44&-78.73&0.1 \\
    SUB 3&40.90&-79.61&0.1 \\
    SUB 4&40.70&-79.26&0.1 \\
    SUB 5&40.70&-79.07&0.1 \\
    SUB 6&41.08&-78.61&0.1 \\
    SUB 7&40.50&-78.20&0.1 \\
    SUB 8&40.53&-78.50&0.1 \\
    SUB 9&41.03&-78.99&0.1 \\
    SUB 10&41.22&-78.35&0.1 \\
    SUB 11&41.48&-79.26&0.1 \\
    SUB 12&41.45&-79.71&0.1 \\
    SUB 13&41.63&-79.75&0.1 \\
    SUB 14&41.86&-79.94&0.1 \\
    SUB 15&42.01&-79.86&0.1 \\
    SUB 16&41.77&-79.45&0.1 \\
    SUB 17&42.01&-78.95&0.1 \\
    SUB 18&41.95&-79.52&0.1 \\
    SUB 19&42.41&-78.73&0.1 \\
    SUB 20&42.02&-78.65&0.1 \\
  \bottomrule
    & \\
    & \\
    \multicolumn{4}{c}{{\footnotesize(c) Other parameters}}\\[0.6ex]
    \toprule
    $\mu$ & \multicolumn{2}{|c}{\$ 1000 /MW (or MVar)} \\[0.6ex]
    $\overline{I}^a_{e}$ &  \multicolumn{2}{|c}{$T_{e}/\min\{\underline{V}_i, \underline{V}_j\}$} \\[0.6ex]
    $r_{e}$ &  \multicolumn{2}{|c}{$(\beta_{e}) r^{o}_{e}$} \\[0.6ex]
    $x_{e}$ &  \multicolumn{2}{|c}{$(\beta_{e}) x^{o}_{e}$} \\[0.6ex]
    $\overline{\theta}$
    &\multicolumn{2}{|c}{30\degree}\\
    \bottomrule 
  \end{tabular}
  }
  \label{tb:para_2}
\end{table}

\vspace{-0.3cm}
\subsection{GIC Modeling Validation}
\begin{table}[htp!]
    \centering
    \footnotesize
    \caption{The test case in \cite{GIC2013flow}, Appendix II}
    \begin{tabular}{ccc}
    \toprule
    Variable Name&Transformer&GIC flow (amps)\\
    \midrule
    $I_{12}$&-&-627.02 \\
    $I_{s}$&T2 series&-763.26 \\
    $I_{34}$&-&-763.26 \\
    $I_{T1}$&T1& 627.02\\
    $I_{c}$&T2 common& 136.24 \\
    $I_{T3}$&T3& -763.26\\
    \bottomrule
    \end{tabular}
    \label{tab:my_label}
\end{table}%
 To validate the GIC modeling in Eq.(\ref{GIC})--(\ref{Idub}), we tested our model on the 6-bus system given in Appendix II of \cite{GIC2013flow} and compared our solution with the results provided in this reference. Table \ref{tab:my_label} displays the GIC flows obtained by solving Eq.(\ref{GIC})--(\ref{Idub}) with fixed $z_{e} = 1$ (in \cite{GIC2013flow} the GIC flows are calculated without line switching options). This solution matches the results found in the reference (see Eq.(B.5)--Eq.(B.10)).

\subsection{Case C1: Potential Damage by GICs}

\begin{figure}[!h]
  \captionsetup{font=footnotesize}
  \centering
  \subfigure[Optimal topology $\mathbf{z}^*_o$ in case C1]{
  \includegraphics[scale=0.29]{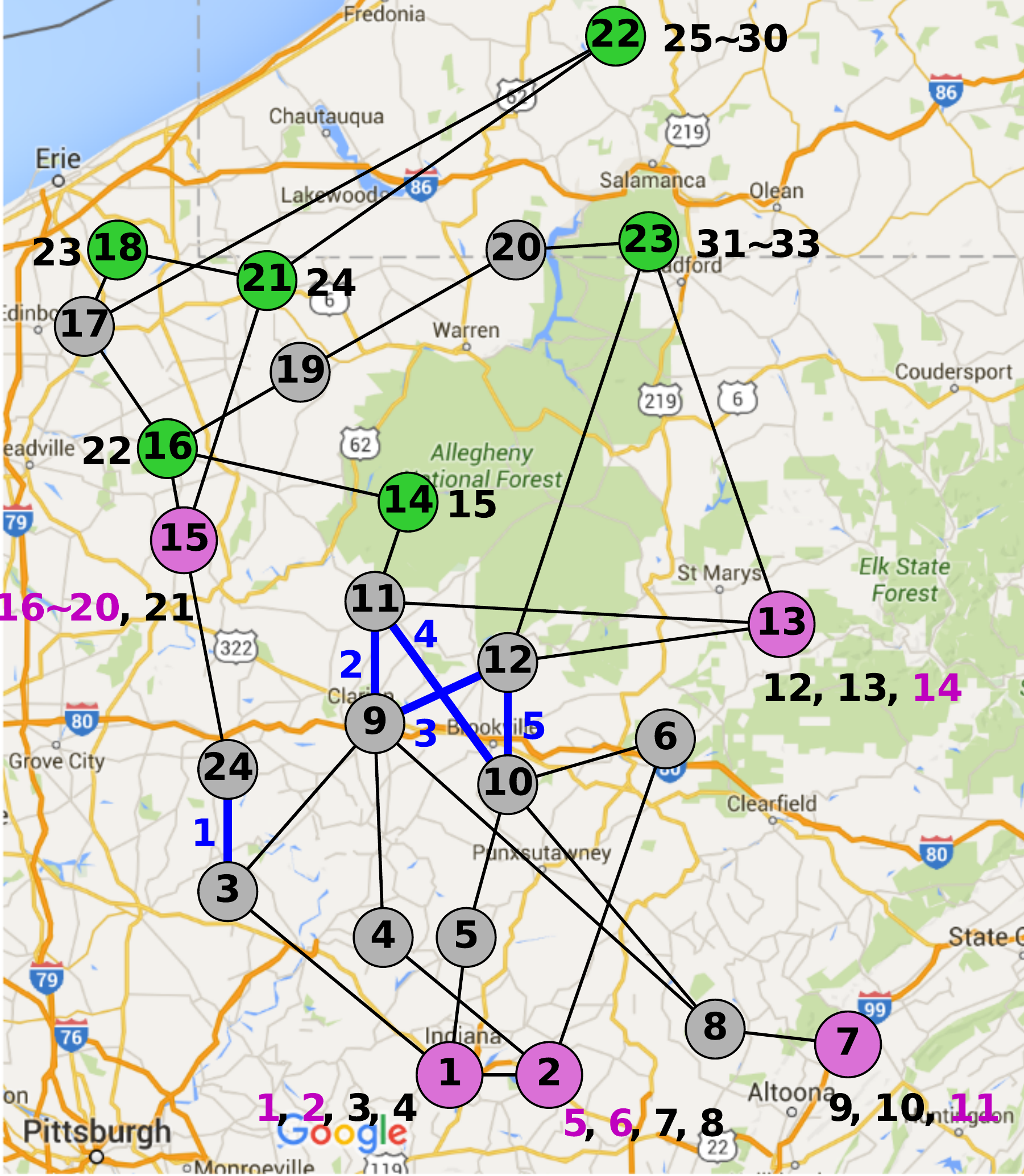}
  \label{fig:normal}
  }
  \subfigure[Potential damage of transformers by GICs]
	{
  \includegraphics[scale=0.355]{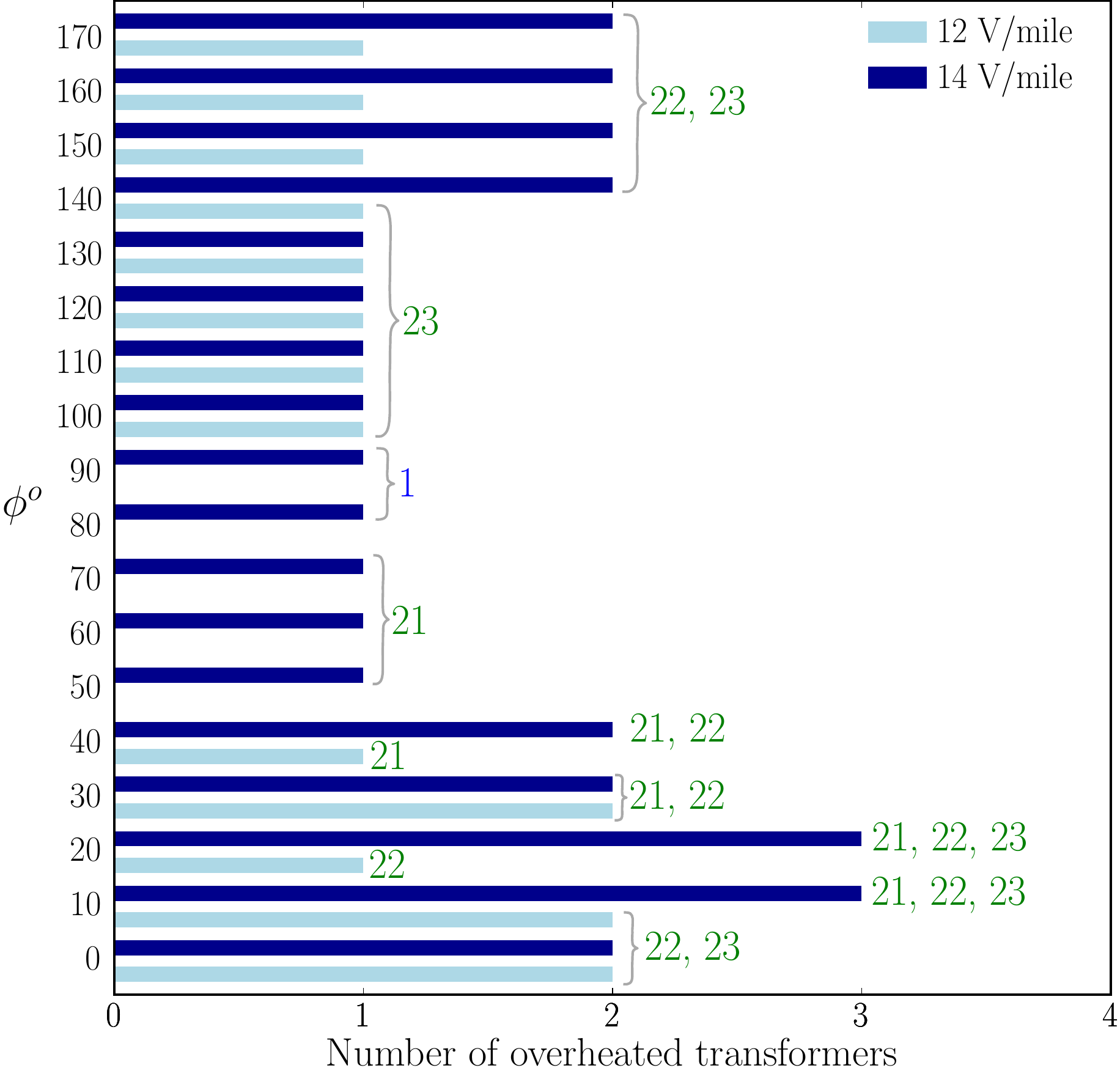}
  \label{fig:damagedtrans}
  }%
\caption{Evaluation of the power system in Table~\ref{tb:para_1}--\ref{tb:para_2} for case C1. (a) The grey nodes are loads. The blue lines indicate network transformers. The green and magenta nodes indicate GSU transformers. Transformer IDs are listed next to the node. The solution to case C1 does not allow reconfiguration of the topology via the network transformers, however, if a generator is not injecting real or reactive power into the network, its GSU is disconnected using the circuit breakers in Fig.~\ref{Fig:4_bus_example}. Generation nodes with disconnected GSU transformers are magenta and their (network or GSU) transformer IDs are marked as magenta as well. (b) The case C1 solution is tested by applying geo-electric fields of strength 12 V/mile and 14 V/mile for all directions. The label above each bar indicates IDs of overheated transformers. Red and green are used to label network transformers and GSU transformers, respectively. 
}
\label{Fig:case1}
\end{figure}

Under normal circumstances without GMDs, line switching decisions are determined by economic dispatch. More specifically, the optimal system topology is obtained by solving an ACOTS model without the GIC-effects constraints
(Case C1). Figure \ref{fig:normal} shows the optimal normal topology, $\mathbf{z}_o^*$, where some generators are not injecting real or reactive power. For example, generators 16 through 20 are shut down at node 15, and their GSU transformers are disconnected from the network using the circuit breakers. Referring to Fig.~\ref{Fig:4_bus_example}, we note that this action does not significantly affect the topology of the AC network, which is only affected by switching transmission lines. This action removes GSU transformer ground points from the DC network topology over which the GICs flow.

Case C1 assumes that generation and system topology are optimized for cost while neglecting the impact of GICs. This impact is calculated using Eqs. (\ref{GIC}) through (\ref{Idmag}) to evaluate the feasibility of thermal limit constraint (\ref{Idub}). Figure \ref{fig:damagedtrans} shows how many GSU and network transformers would be overheated under C1 depending on the direction and strength of the GMD. Figure~\ref{fig:damagedtrans}, presents results for $c^*_{gmd}$ from 0 to 180$^\circ$ because the strength of the geo-electric field is uniform, and the effects do not depend on field direction. For example, when the electric field is 12 V/mile, the GSU transformer 23 (at node 18) is overheated when the event is oriented between 100$\degree$ and 170$\degree$. When the strength is increased to 14 V/mile, one or more transformers are overheated at almost all orientations of the GMD. For example, when the event is oriented at 10$\degree$,  GSU transformers 21, 22 and 23 are overheated. As the event is shifted to 80\degree,  then network transformer 1 is the transformer at risk. These results provide a baseline to evaluate alternative operating paradigms that ensure system security.

\subsection{Case C2: GIC Mitigation via ACOTS}

Using case C2, the cost benefits of simultaneous controlling generation dispatch and network topology to mitigate GIC effects are evaluated.  
\subsubsection{Cost Analysis}
\begin{figure}[htp]
  \captionsetup{font=footnotesize}
  \centering
  \includegraphics[scale=1.0]{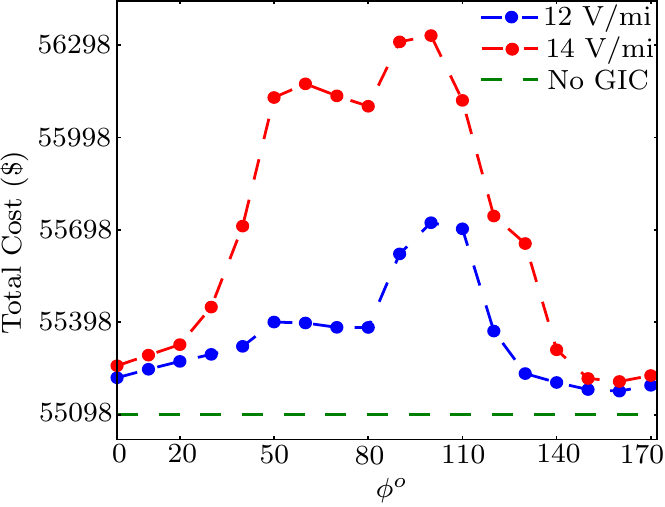}
  \caption{The total cost $c^*_{gmd}$ for case C2 for different geo-electric field orientations and strengths.}
  \label{fig:obj_all}
\end{figure}

For geo-electric field strengths of 12 V/mile and 14 V/mile, case C2 is solved for orientations of the field from 0 to 360$^\circ$, which results in a total cost $c^*_{gmd}$ (see Fig.~\ref{fig:obj_all}) and topology $y^*_{gmd}$ (discussed later). Figure~\ref{fig:obj_all} only presents results for $c^*_{gmd}$ from 0 to 180$^\circ$ because of the symmetry discussed above. 

The results in Fig.~\ref{fig:obj_all} show that the directions of the geo-electric field are not all equivalent because the cost of mitigation $c^*_{gmd}$ varies with direction.  The most costly GMDs occur when the event is oriented between 20$\degree$ and 140$\degree$. The increase in cost between 12 V/mile and 14 V/mile is primarily due to changes in generator dispatch and is not significant. For example, the difference in cost between the 12V and 14V per mile case is 1.40\% when the GMD is oriented at 60$\degree$. {\color{black} Moreover, the dispatch cost is smaller when GIC effects are neglected (Case C1). However, the transformer thermal limit constraints are violated when GIC effects are applied to the network (as seen in Fig. \ref{fig:damagedtrans}). 
Thus, there is an implicit higher cost associated with replacing the damaged equipment and unexpected load shed when the transformer fails.} 

\subsubsection{Topology Control Analysis}
\begin{figure}[htp]
  \captionsetup{font=footnotesize}
  \centering
  \subfigure[12 V/mile, 80\degree]{
  \includegraphics[scale=0.182]{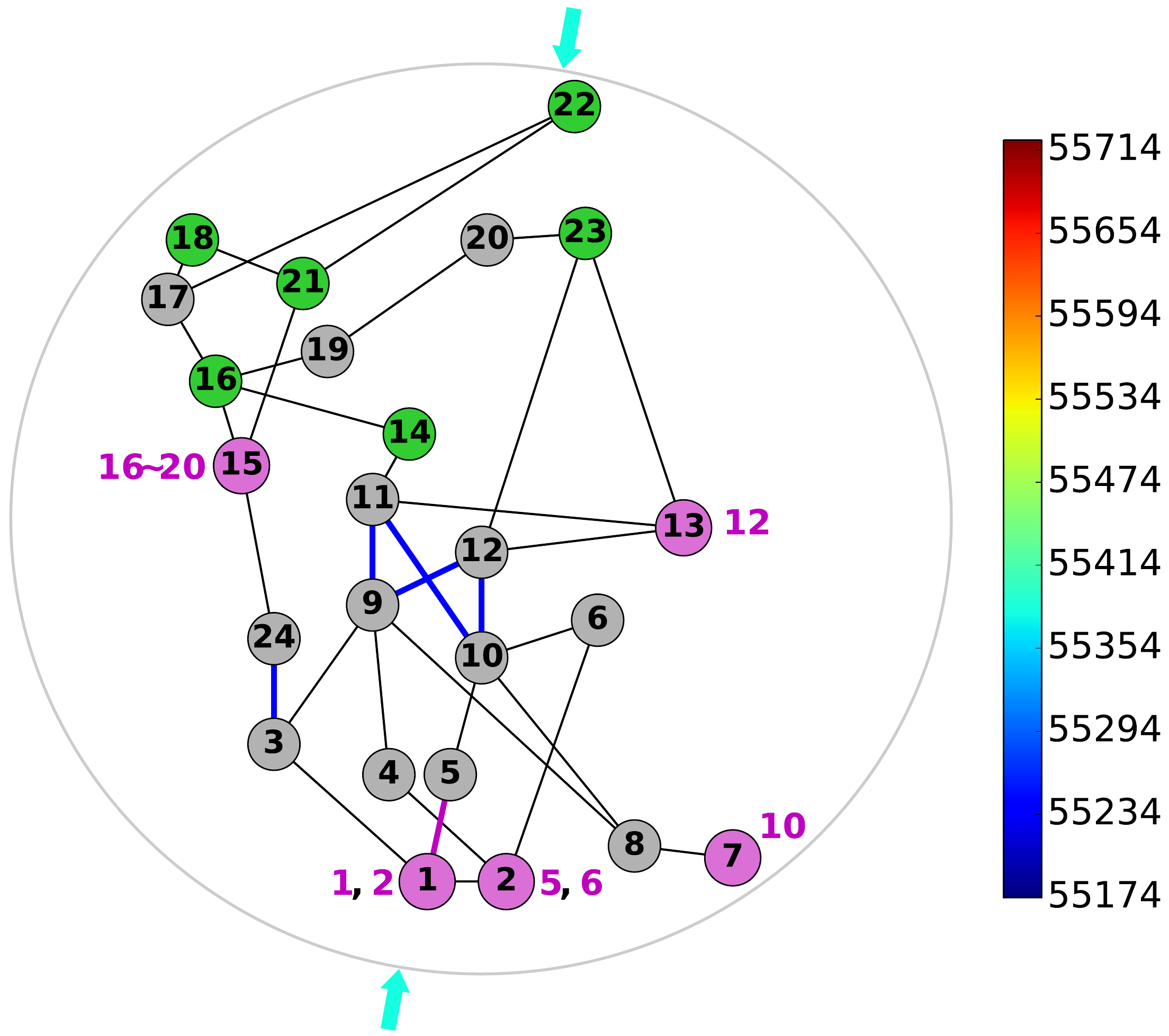}
  \label{fig:5_40_sunplot}
  }
  \subfigure[12 V/mile, 90\degree]{
  \includegraphics[scale=0.182]{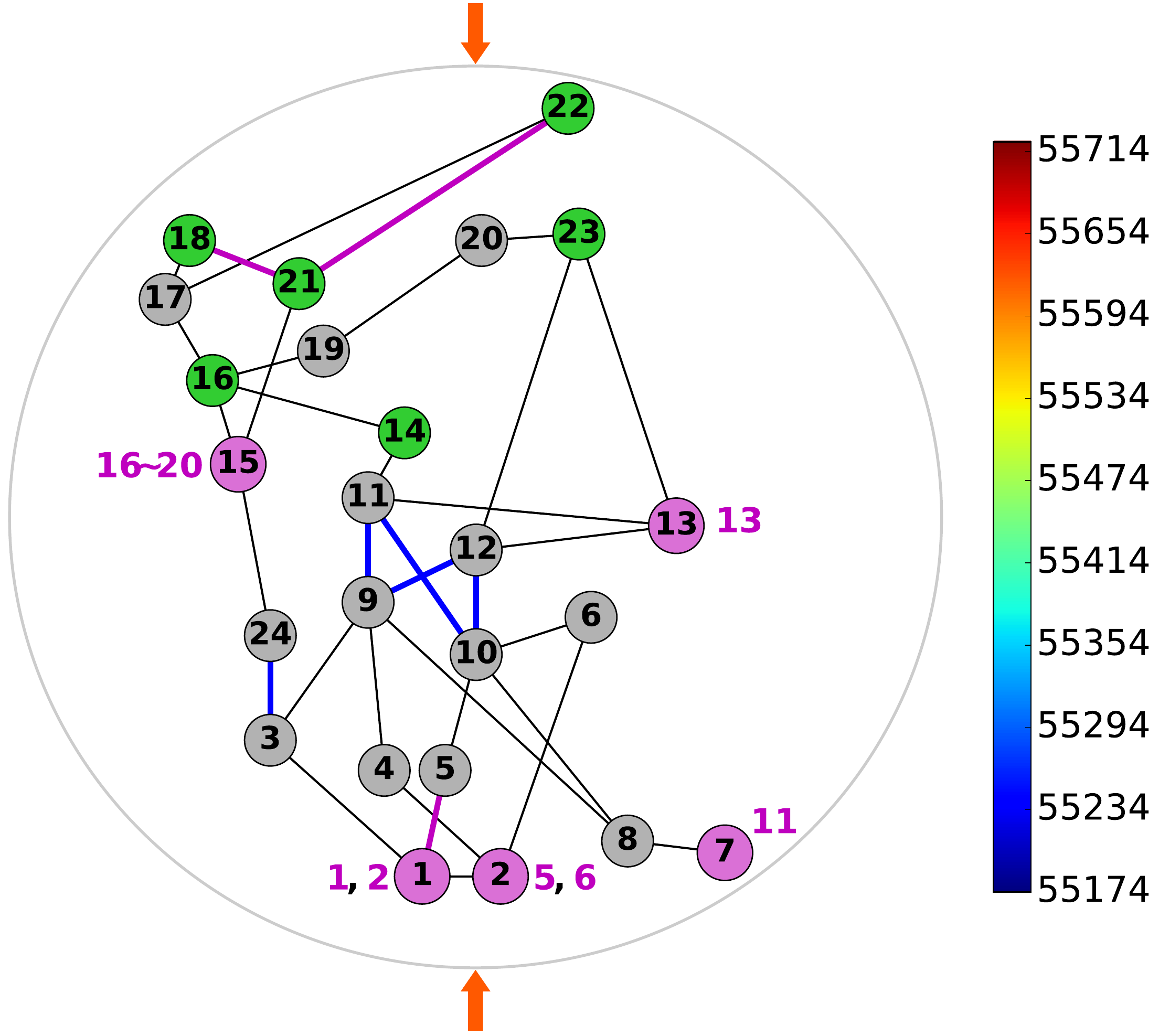}
  \label{fig:5_50_sunplot}
  }
  \subfigure[12 V/mile, 100\degree]{
  \includegraphics[scale=0.182]{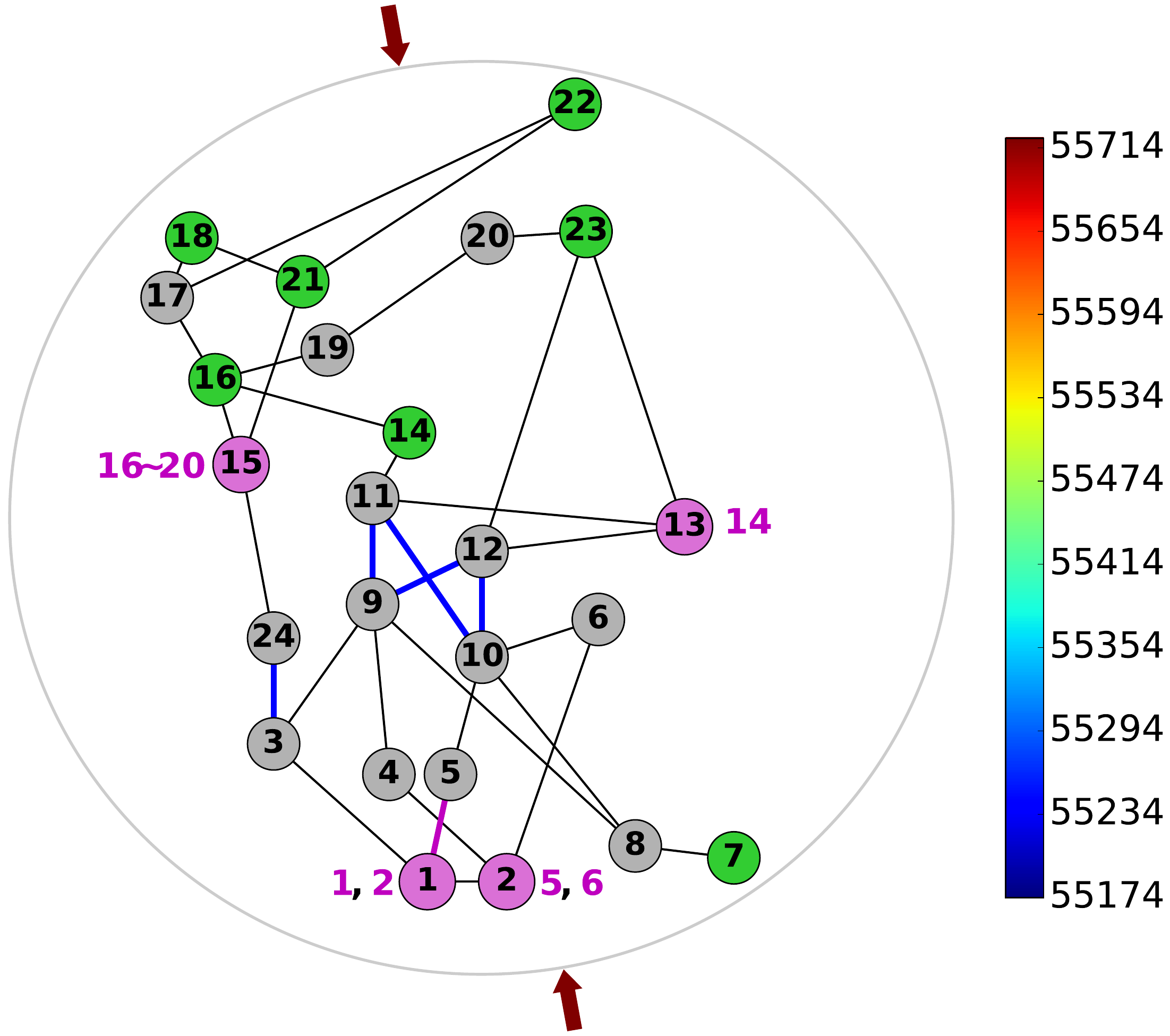}
  \label{fig:5_60_sunplot}
  }
  \subfigure[12 V/mile, 110\degree]{
  \includegraphics[scale=0.182]{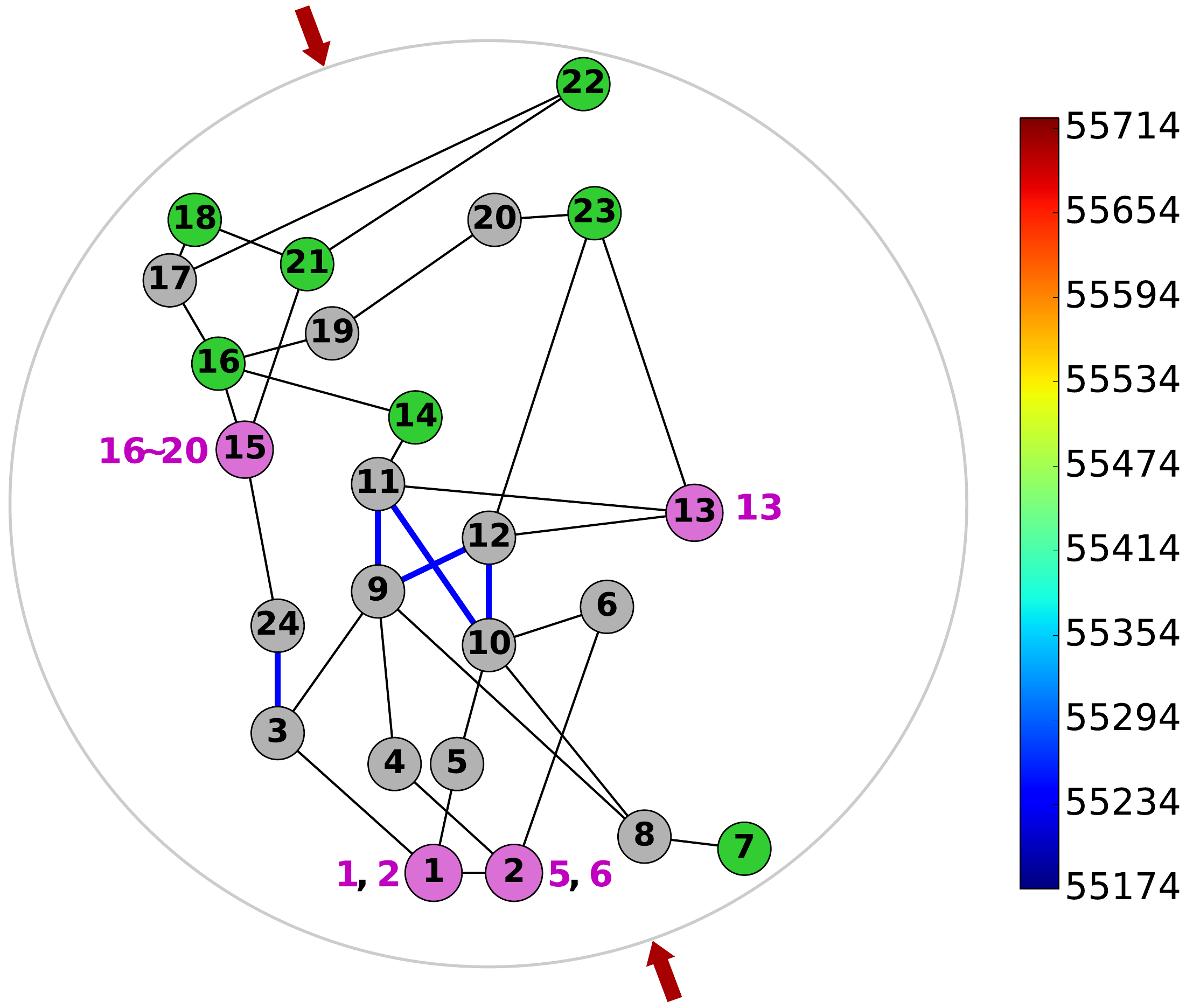}
  \label{fig:5_70_sunplot}
  }
  \caption{Topology solutions for case C2 at 12 V/mile strength and orientations from 80$\degree$--110$\degree$. Switched off lines are colored magenta and the IDs of unused generators are labeled beside their connected substations. 
  }
  \label{fig:5_sens_sunplot}
  \vspace{-0.45cm}
\end{figure}

\begin{figure}[htp]
 \captionsetup{font=footnotesize}
  \centering
  \subfigure[14 V/mile, 80\degree]{
  \includegraphics[scale=0.182]{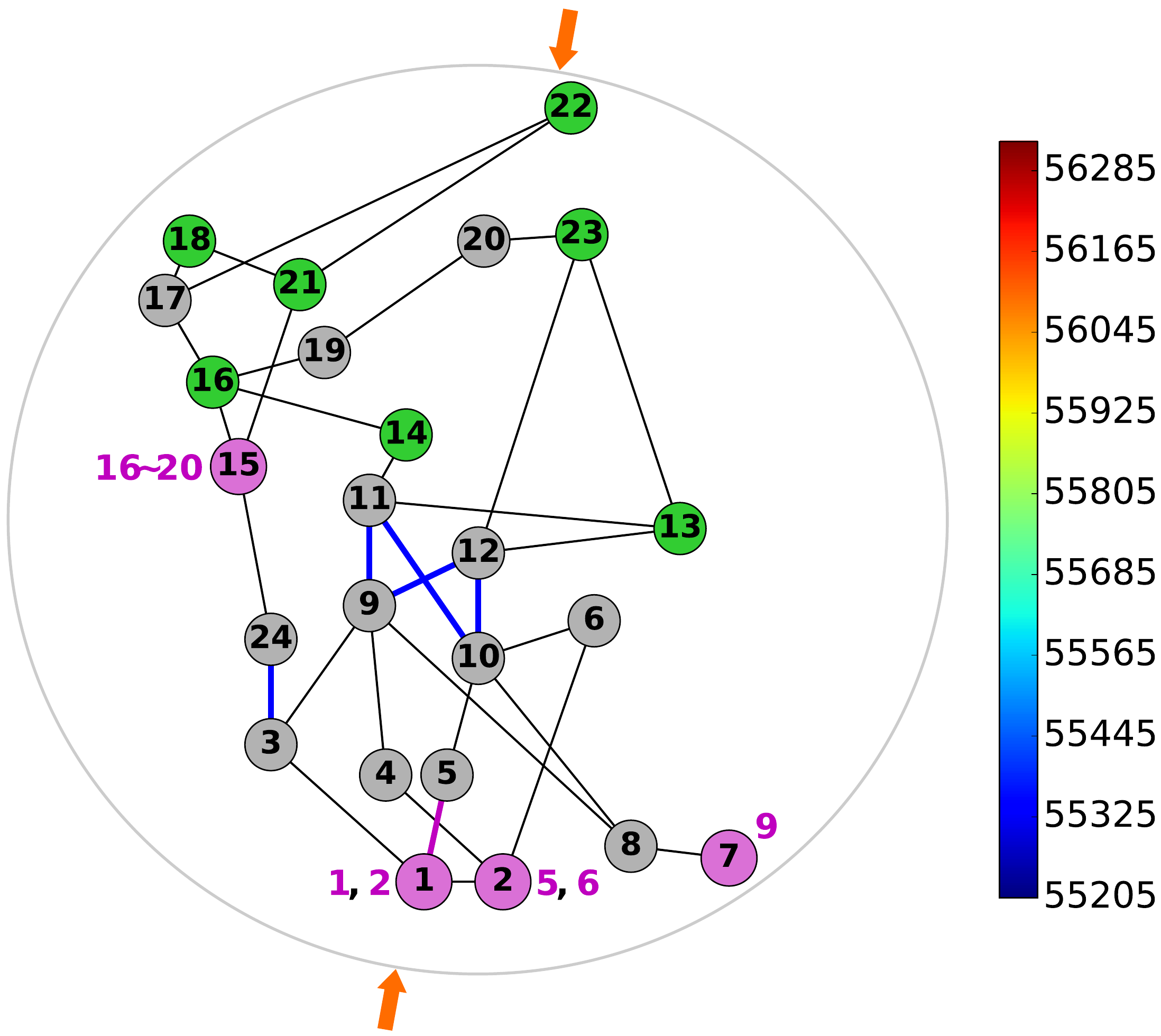}
  \label{fig:7_40_sunplot}
  }
  \subfigure[14 V/mile, 90\degree]{
  \includegraphics[scale=0.182]{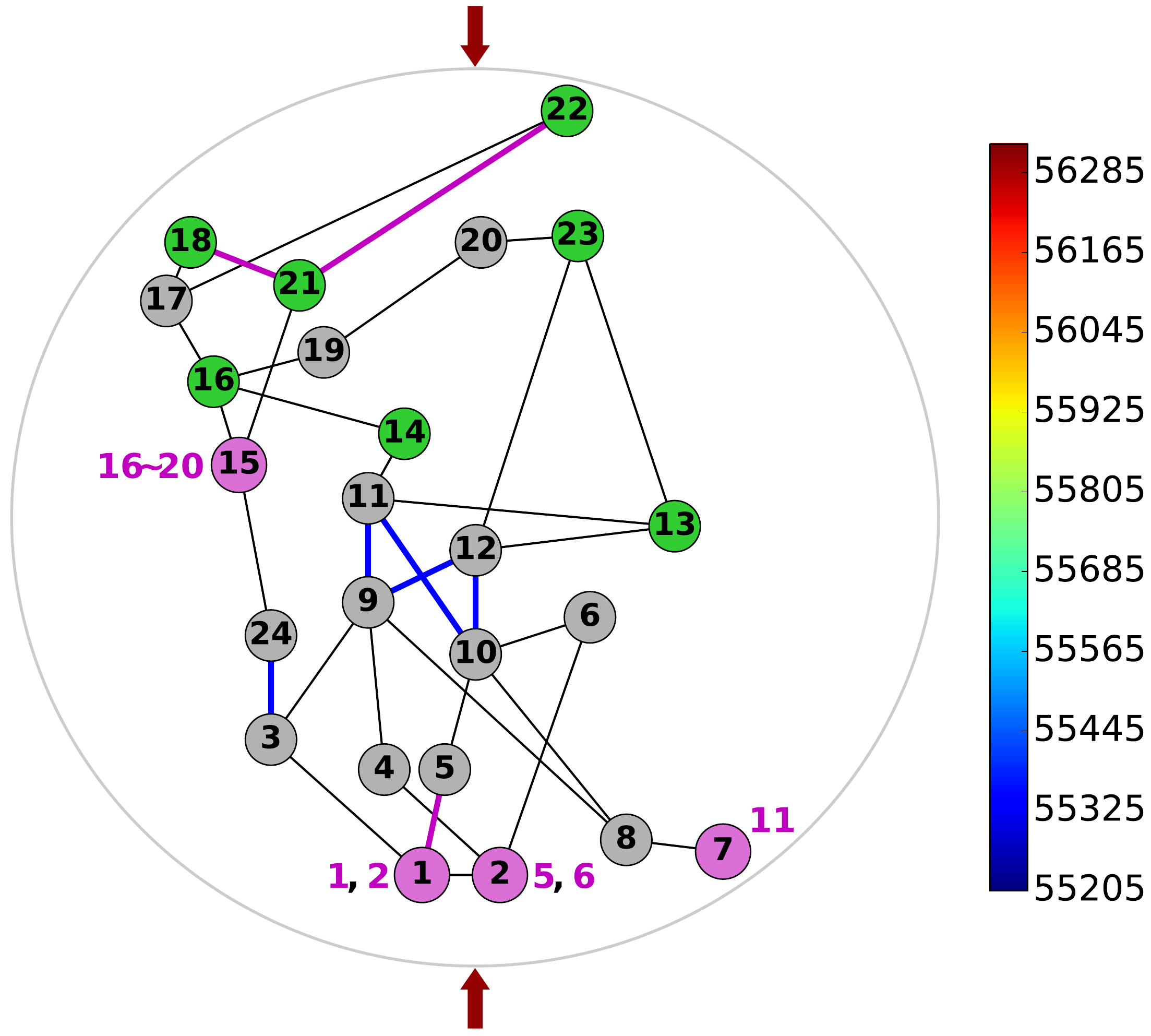}
  \label{fig:7_50_sunplot}
  }
  \subfigure[14 V/mile, 100\degree]{
  \includegraphics[scale=0.182]{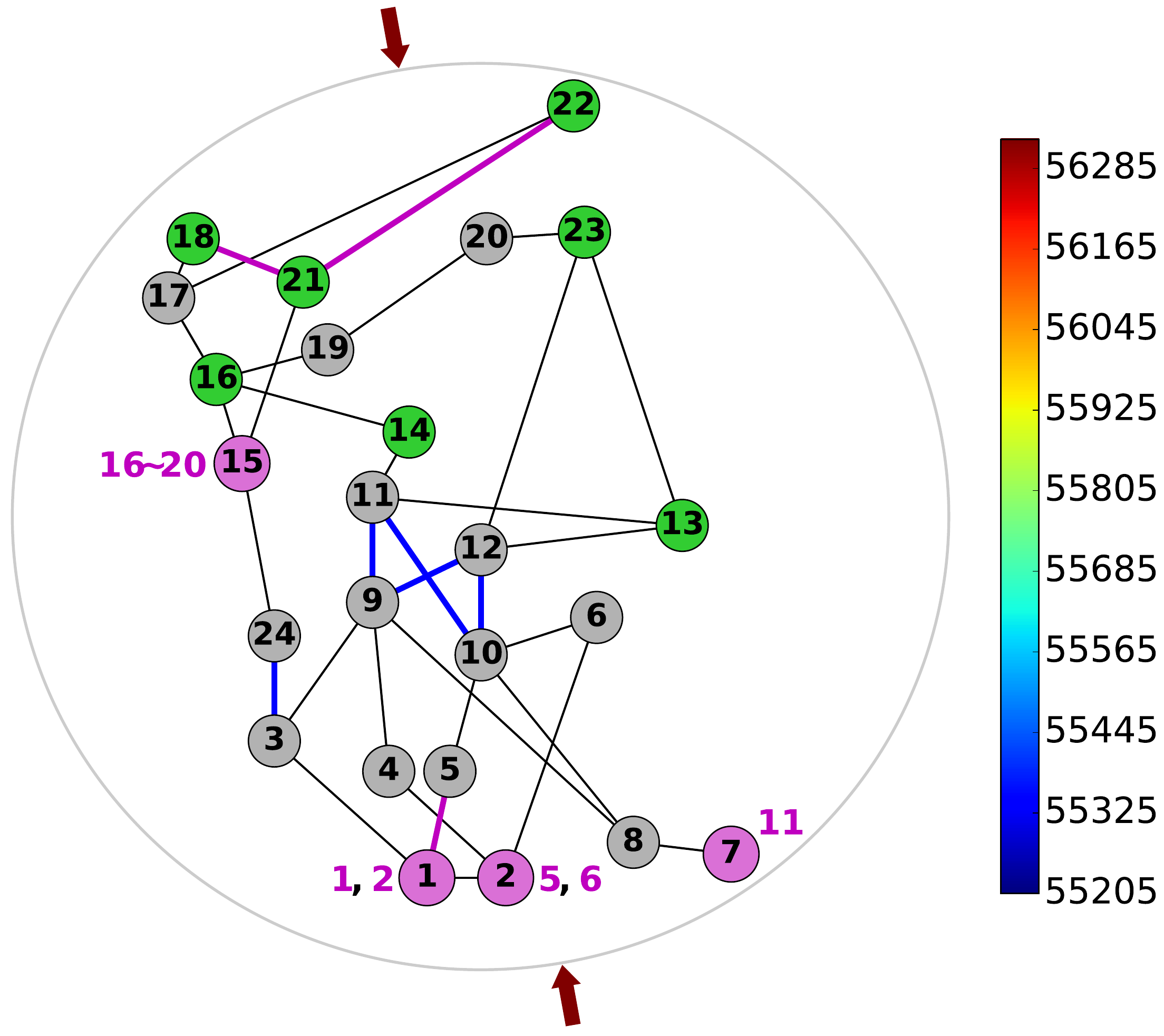}
  \label{fig:7_60_sunplot}
  }
  \subfigure[14 V/mile, 110\degree]{
  \includegraphics[scale=0.182]{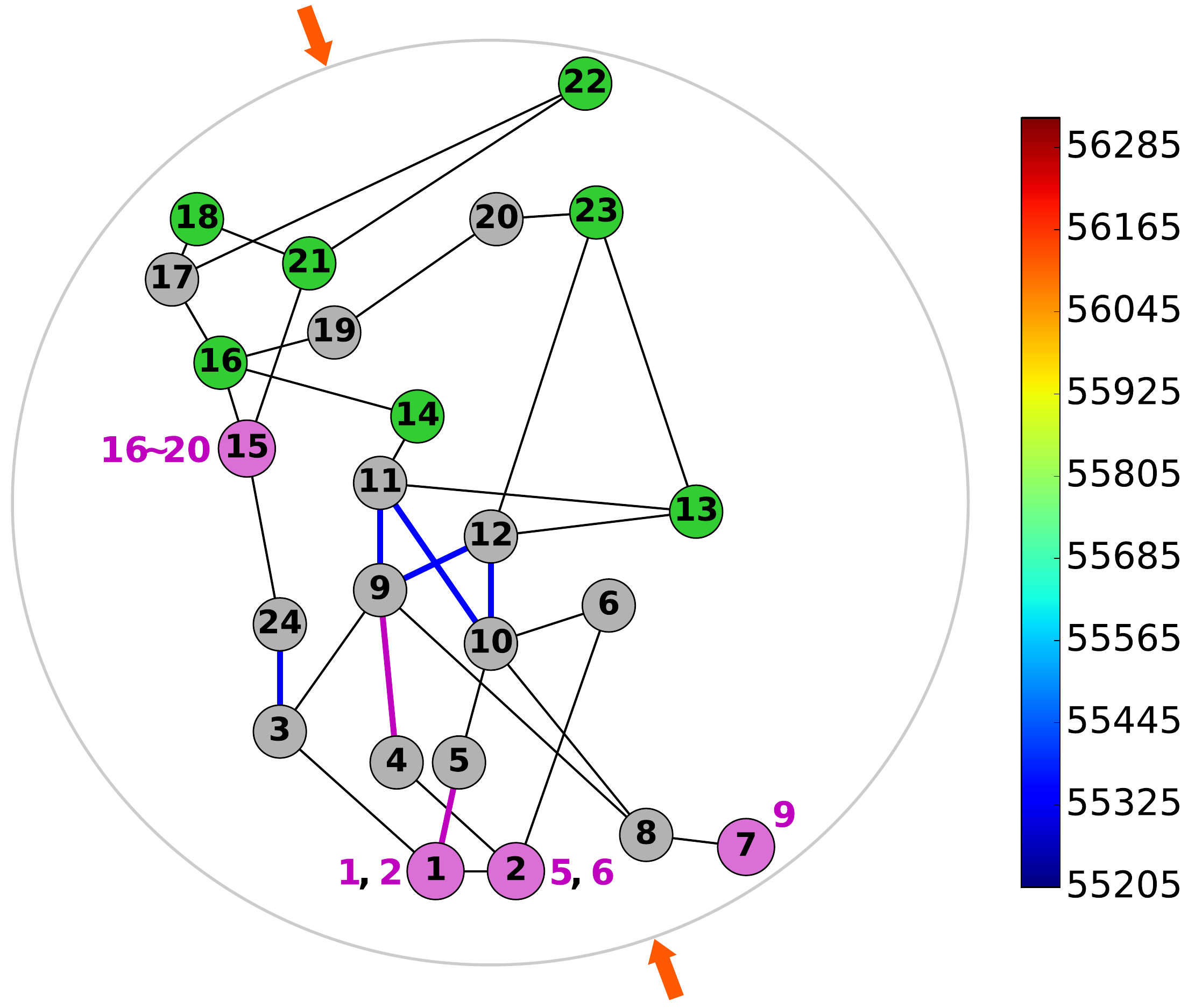}
  \label{fig:7_70_sunplot}
  }
  \caption{Same as Fig~\ref{fig:5_sens_sunplot} but for a geo-electric field strength of 14 V/mile.  
  }
  \label{fig:7_sens_sunplot}
\end{figure}

In Fig.~\ref{fig:obj_all}, the topology of the network varies with the strength and direction of the GMD. In the remainder of this section, we focus on GMD events oriented between 80$\degree$ and 110$\degree$ because these directions are most sensitive to GMD. Figures \ref{fig:5_sens_sunplot} and \ref{fig:7_sens_sunplot} display the network topology for geo-electric field strengths of 12 V/mile and 14 V/mile, respectively.\footnote{We note that there are multiple generators located at buses 1, 2 and 15.  Generators 1 and 2 (at bus 1) have the same cost and capacity, as do generators 5 and 6 at bus 2 and generators 16 through 20 at bus 15. Thus, there are equivalent dispatch solutions.}

In Fig.~\ref{fig:5_sens_sunplot}, in the 80$\degree$ geo-electric field case, only one transmission line is switched off (1,5). In the 90$\degree$ geo-electric field case, three transmission lines are switched off. Two are intuitively long lines oriented along the geo-electric field and incur larger GICs. Switching these lines off removes a significant vulnerability. The third line is nearly perpendicular to the geo-electric field and is also switched off. This counter-intuitive topology control is being used to reroute power flow away from other, more susceptible transmission lines. 

In the 100$\degree$ orientation case, some of the lines in the 90$\degree$ case remain in the solution and some disappear. All transmission lines are switched on when the event orientation is at 110$\degree$. Thus, the sensitivity of the topology solution to the details of the orientation and the difficulty in making accurate predictions of geo-electric field direction suggest that the ACOTS formulation should be extended to a stochastic formulation over the field direction in future work.

The results displayed in Fig.~\ref{fig:7_sens_sunplot} for different geo-electric field orientations suggests similar conclusions. At a fixed 14 V/mile in Fig.~\ref{fig:7_sens_sunplot}, the optimal topology solutions switch off several long transmission lines oriented along the geo-electric field, but some transmission lines still display significant sensitivity to orientation.  

Comparing Fig.~\ref{fig:7_sens_sunplot} (14 V/mile) with Fig.~\ref{fig:5_sens_sunplot} (12 V/mile) shows that some topology solutions at low field strength persist to higher field strength, however significantly more transmission lines are switched off to avoid large GIC in the network and in GSU and network transformers. The properties of the topology solutions for different geo-electric field strengths again suggests that the ACOTS solution should be extended to a stochastic or robust formulation over field strength.  
Finally, we note that while the solution adjusts the topology, it does not create islands---a mitigation strategy that is sometimes suggested. However, islands could form in larger, more complex networks. 

%

\vspace{-0.1cm}
\subsection{Case C2 versus Case C3: Cost Benefits of Topology Optimization}

\begin{figure}[htp]
    \captionsetup{font=footnotesize}
    \centering
    \includegraphics[scale=0.35]{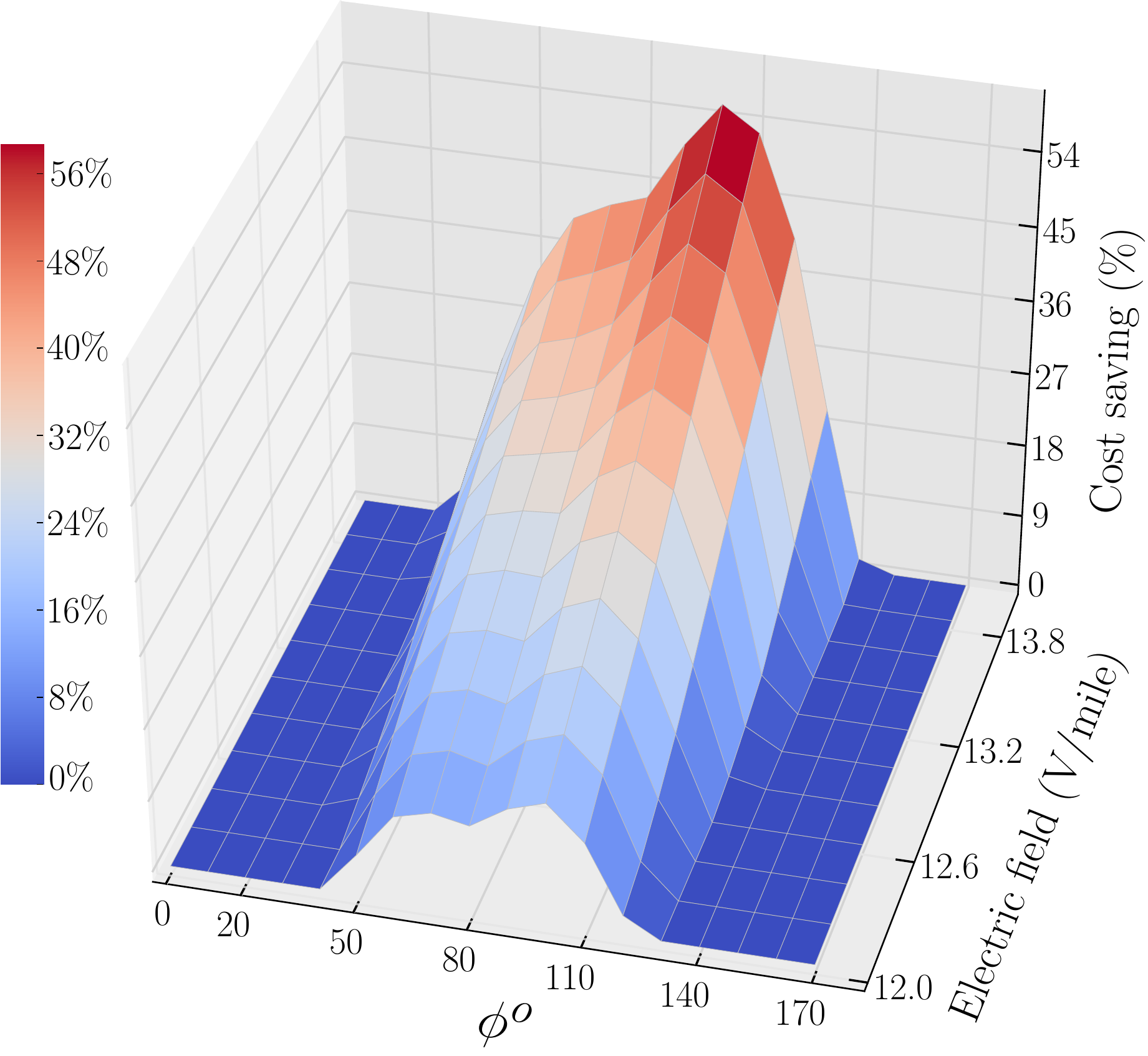}
    \caption{Combined savings from generator dispatch and load shedding costs enabled by the optimal topology $\mathbf{z}^*_{gmd}$ found by ACOTS relative to the dispatch and load shedding cost incurred by the ACOPF of case C3 with the topology fixed to $\mathbf{z}^*_{0}$.}
    \label{fig:3d}
    \vspace{-0.3cm} 
\end{figure}

The inclusion of topology control into the ACOTS formulation increases the complexity of the problem, but it also provides significant cost savings over a less complex ACOPF. The cost savings is evaluated by comparing case C2 (where topology control is allowed) with case C3 (where the topology is fixed to that found in case C1). Figure \ref{fig:3d} displays the percentage cost savings of C2 (ACOTS) over C3 (ACOPF) for field strengths between 12 and 14 V/mile and field directions between 0$\degree$--180$\degree$. Under the most severe GMD conditions explored, the benefit of topology control is as much as 54\%.

Table \ref{table:perc_cost} further breaks down the cost savings of case C2 over case C3 into generator dispatch costs and load shedding costs. For the 14 V/mile field strength case, the topology control in case C2 enables nearly all of the load to be served.  In contrast, the fixed topology in C3 results in load shedding costs of 13.9\% on average and 33.54\% in the worst case. 

\begin{table}[htp]
  \captionsetup{font=footnotesize}
  \captionof{table}{Percentage of the total cost in cases C2 and C3 due to generator dispatch and load shedding. For 12 V/mile and 14 V/mile strengths, the average, minimum and maximum percentage of total cost is computed over the geo-electric field orientation from 0$\degree$--180$\degree$.} \label{tab:title} 
  \begin{tabular}{lrrrrrrr}
  \toprule
       &&\multicolumn{3}{r}{Dispatch Cost(\%)}&\multicolumn{3}{r}{Load shedding Cost(\%)}\\
      \cmidrule(lr){3-5}
      \cmidrule(lr){6-8}
       Strength&Case&Avg.&Min.&Max.&Avg.&Min.&Max.\\ 
       \midrule
       \multirow{ 2}{*}{12 V/mile}
       &C2&100.0&100.0&100.0&0.0&0.0&0.0 \\
       &C3&97.8&90.8&100.0&2.1&0.0&9.2 \\
       \midrule
       \multirow{ 2}{*}{14 V/mile}
       &C2&100.0&100.0&100.0&0.0&0.0&0.0 \\
       &C3&86.1&66.46&100.0&13.9&0.0&33.5 \\
       \bottomrule
 \end{tabular}\par
 \label{table:perc_cost}
\end{table}%

\subsection{Case C3 versus Case C4: Performance of Network Reconfiguration}
\begin{figure}[htp]
  \captionsetup{font=footnotesize}
  \centering
  \subfigure[The total cost $c^*_{f}$ in Case C3.]
	{
  \includegraphics[scale=0.145]{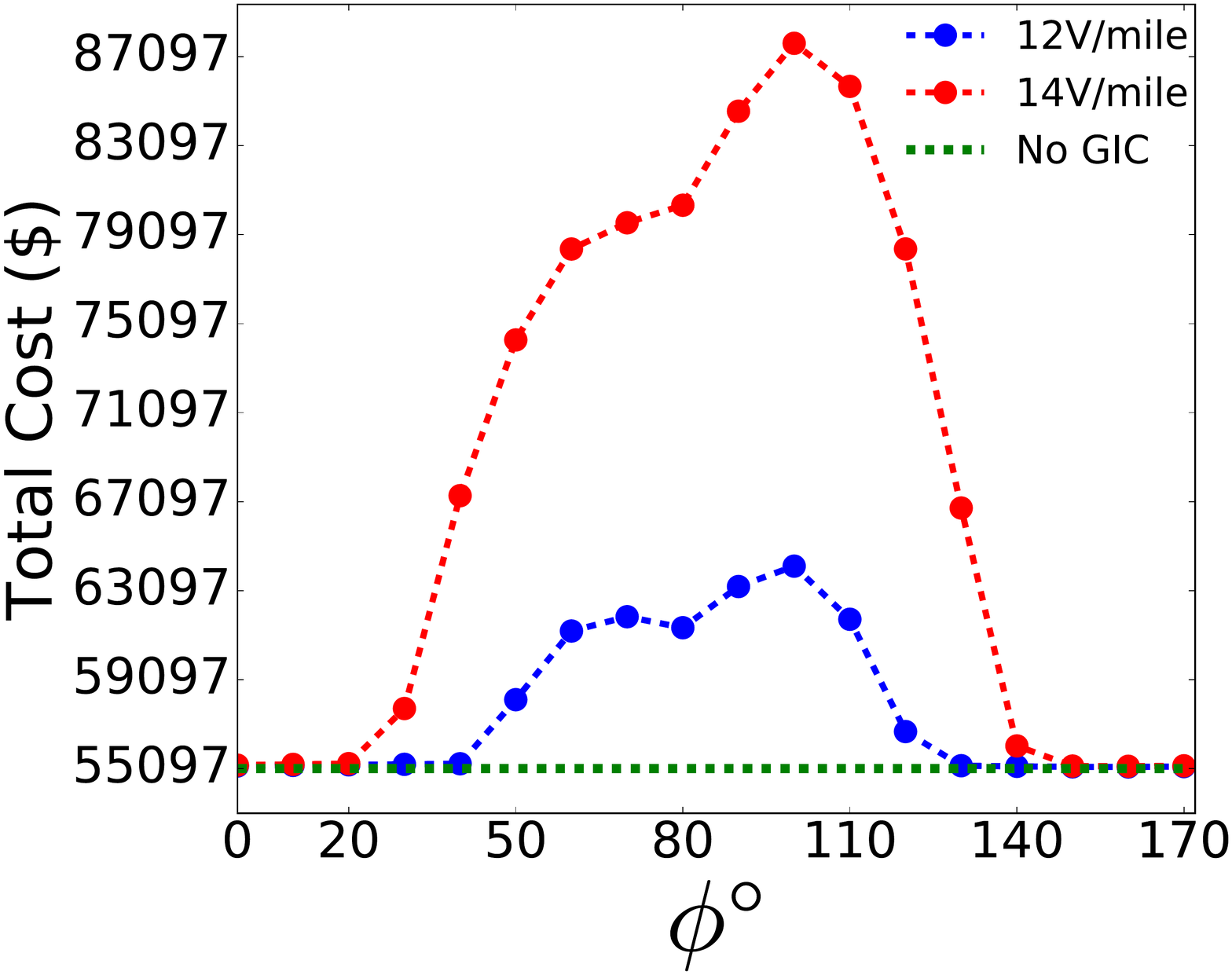}
  \label{fig:c3}
  }
  \subfigure[The total cost $c^*_{\mathbf{1}}$ in Case C4.]
	{
  \includegraphics[scale=0.145]{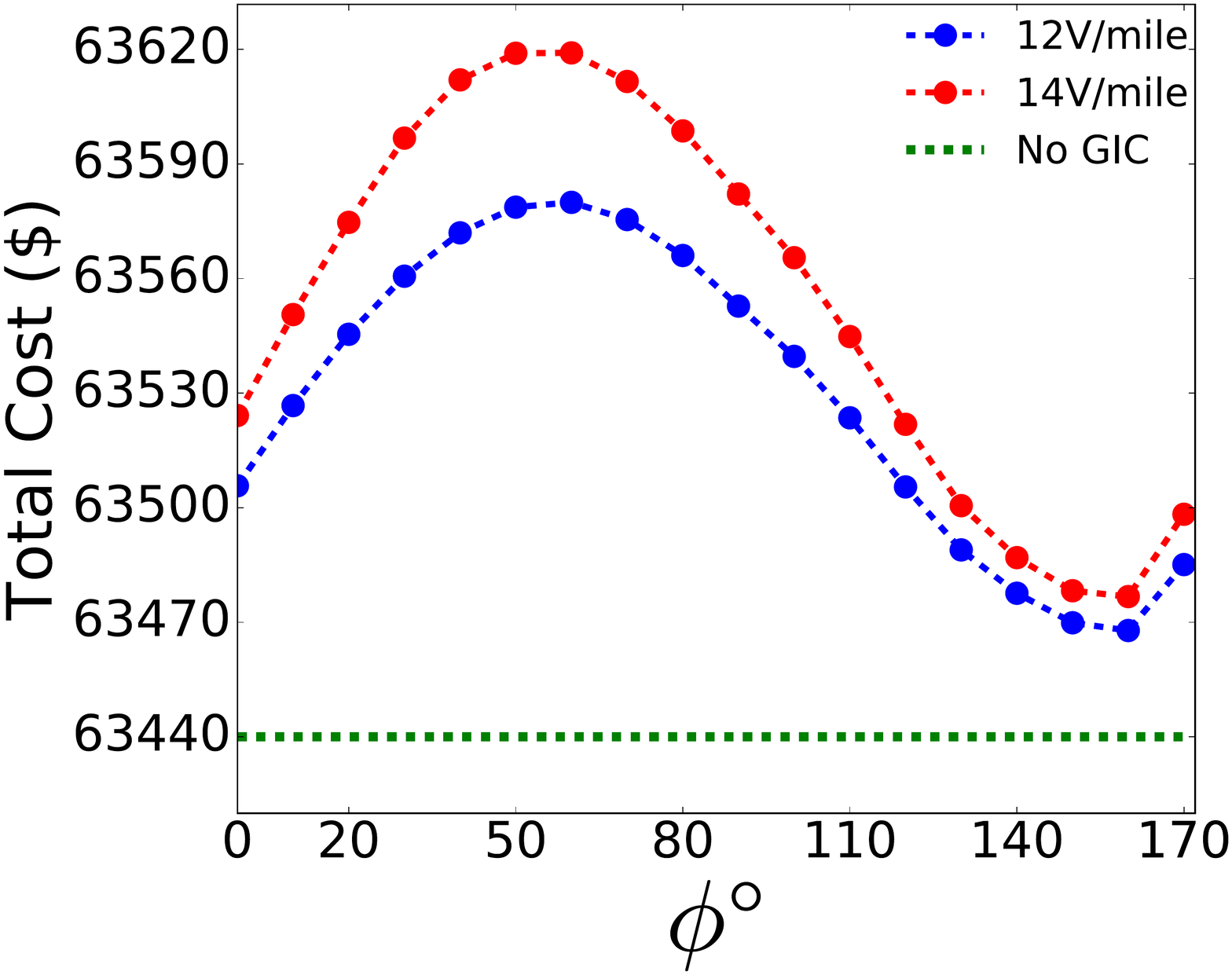}
  \label{fig:c4}
  }%
\caption{Cost comparison of Case C3 and C4 for different geo-electric field orientations and strengths.
}
\label{Fig:c3_vs_c4}
\vspace{-0.2cm}
\end{figure}%

The results displayed in Figure \ref{Fig:c3_vs_c4} compare the topology of case C3 (where  topology  is  fixed  to case  C1) with case C4 (where all lines and generators). {\color{black} Similar to the results in Fig. \ref{fig:c3}, the total cost, $c^*_{\mathbf{1}}$, varies with event direction. Ignoring GIC effects also induces a lower operating cost (Fig. \ref{fig:c4}). However, like the results of Fig. \ref{fig:damagedtrans}, we also observed that the transformer thermal limit constraints (\ref{thermal}) are violated when GIC effects are neglected. For example, if the field strength is 14 V/mile and has an orientation of $20\degree$, GSU transformers 22 and 23 overheat.} Comparing Fig. \ref{fig:c3} (C3) with Fig. \ref{fig:c4} (C4) shows that the optimal topology control found in case C1 induces a higher cost than case C4 for orientations through 40$\degree$ to 130$\degree$ under the field of 14 V/mile. This is due to forced disconnect of generators in Case C1 (e.g., generator 14 at bus 13) which could be dispatched more effectively, when no line can be switched off, to mitigate GIC effects.

\subsection{Computational Analysis}

\subsubsection{Computational Speed} Table~\ref{table:walltime} summarizes the computational time properties of ACOTS with GIC subject to convex relaxations (from section \ref{subsec:ConvexRelaxation}) by solving them to optimality. We observed that the times are higher at larger geo-electric field strength, likely because of the increased complexity due to larger number of possible topology changes. Though the computations are time-intensive (in Table~\ref{table:walltime}), we observed that by terminating the solver at larger optimality gaps (say 5\%), a solution was obtained within 320 seconds, which was 2-3 orders of magnitude quicker than solving to optimality.

\begin{table}[htp]
    \captionsetup{font=footnotesize}
    \centering
    \caption{Computational time for geo-electric field strengths of 12 V/mile and 14 V/mile {\color{black} on the RTS system}. The average, minimum, maximum and standard deviation of solving time are presented over the geo-electric field orientation from 0$\degree$--360$\degree$. {\color{black}Results without parentheses present elapsed time for solving to optimality. Solutions displayed in parentheses denote solving time when the optimization is terminated with a 5\% optimality gap.} }
    \begin{tabular}{crrrr}
    \toprule
         &\multicolumn{4}{c}{Wall Time (sec)}\\ 
         \cmidrule{2-5}
         Strength(V/mile)&Avg.&Min.&Max.&Std. dev.\\
    \midrule
        \multirow{2}{*}{12}&167.1 &31.5 &1066.9 &264.1 \\
        &\color{black}(10.6) & \color{black}(10.2) & \color{black}(11.0) & \color{black}(0.3)\\
        \multirow{2}{*}{14}&1249.9&32.6&6676.3&1880.9 \\
        &\color{black}(56.0) & \color{black}(10.3) & \color{black}(314.1) & \color{black}(85.8)\\
    \bottomrule
    \end{tabular}
    \smallskip
    \label{table:walltime}
    \vspace{-0.3cm}
\end{table}%

\subsubsection{Scalability to Larger Network} 
\begin{figure}[htp]
  \captionsetup{font=footnotesize}
  \centering
  \includegraphics[scale=0.39]{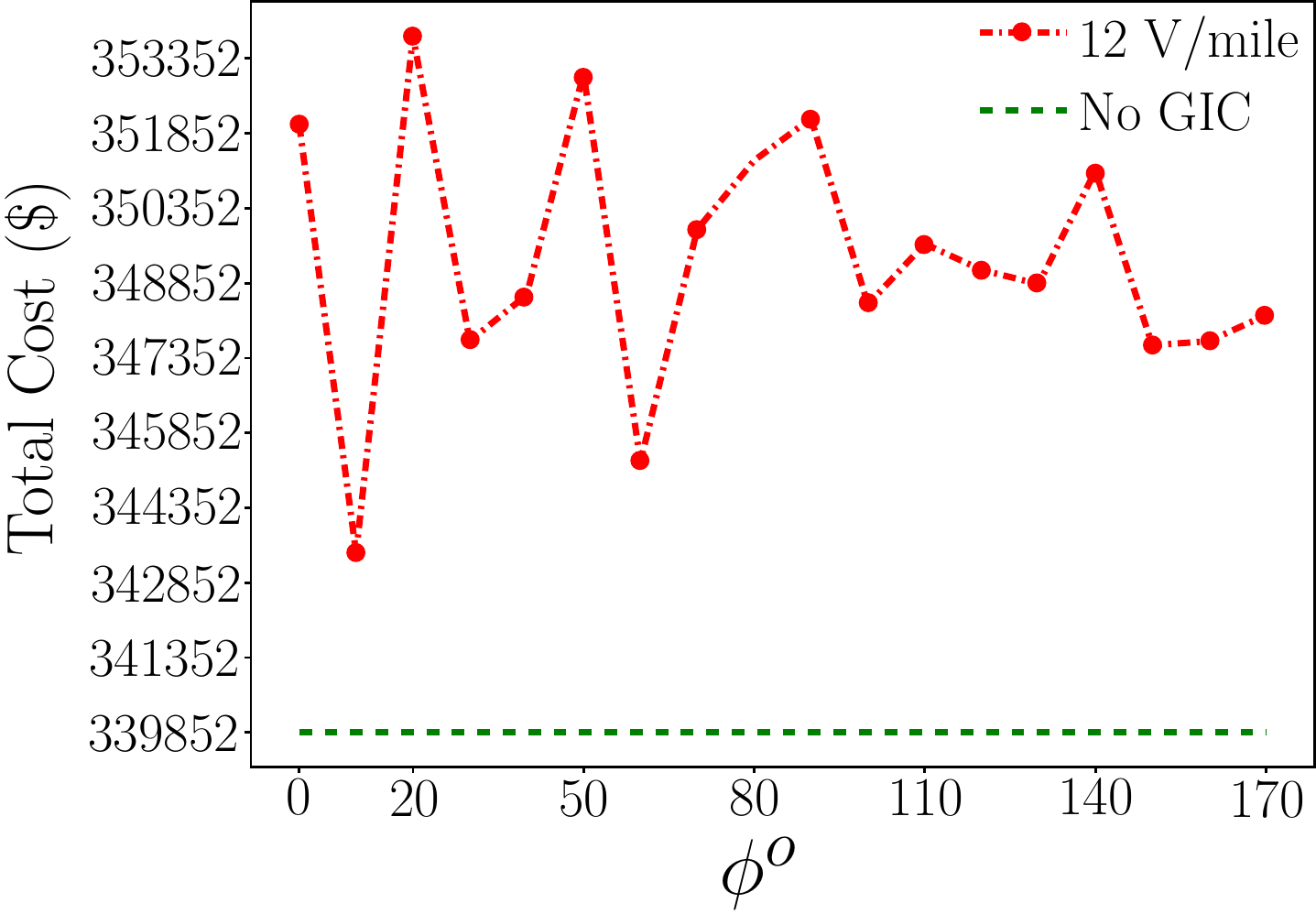}
  \caption{\color{black} Feasible solution costs of UIUC 150-bus system for geo-electric field strength of 12 V/mile in Case \ref{case2}.}
  \label{fig:obj_150}
   \vspace{-0.4cm}
\end{figure}

The computational time required to solve small test cases implies that a key limitation for practical deployment of this model is scalability. This is not surprising as solution methods for OTS suffer similar limitations. However, in the case of GMD mitigation, high-quality solutions that are close to optimality are often sufficient. {\color{black} On \textit{UIUC 150-bus system}, figure \ref{fig:obj_150} shows the feasible solution costs when the optimization is terminated with a 5$\%$ optimality gap. Like are other results, the cost of dispatch varies with direction and is higher than the dispatch cost when GIC is ignored. In Table \ref{tb:uiuc_150}, we provide the computational times required to obtain results displayed in Fig. \ref{fig:obj_150}}. These results suggest that it is practical to use heuristic methods in conjunction with \textit{state-of-the-art} convex relaxations to find high-quality solutions to larger scale systems on the time scales required for GMD mitigation efforts. 

\begin{table}[htp]
    \captionsetup{font=footnotesize}
    \centering
    \caption{Computational time of the UIUC 150-bus system. The average, minimum, maximum and standard deviation of solving time are presented over the geo-electric field orientation from 0$\degree$--360$\degree$.}
    \begin{tabular}{crrrr}
    \toprule
         &\multicolumn{4}{c}{Wall Time (sec)}\\ 
         \cmidrule{2-5}
         Strength(V/mile)&Avg.&Min.&Max.&Std. dev.\\
    \midrule
        12&996.3&648.9&1586.6&290.4 \\
        14&1162.3&508.1&2896.1&750.2 \\
    \bottomrule
    \end{tabular}
    \smallskip
    \label{tb:uiuc_150}
    \vspace{-0.25cm}
\end{table}%

\subsection{Recovering AC Feasible Solutions}\label{sec:recover} 
The ACOTS with GMD constraints formulated in this manuscript is solved applying several hierarchical convex relaxations of the AC power flow physics. Since the solutions obtained may not necessarily lie in the original nonconvex feasible region, we present a simple methodology to test the quality of the relaxed solutions and obtain AC feasible solutions. 

In \cite{hijazi2013convex}, for OPF-based problems, it is empirically shown that the lower bounds obtained from convex quadratic relaxations are mostly close to globally optimal objective values. Thus, we exploit this fact and apply an objective-cost-based constraint to the problem of ACOTS with GMD. Let the optimal value of formulation  \eqref{eq:ACGMD} with convex relaxations be $\mathcal{O}^*_{lb}$ and let $z^*_e \ \forall e \in \mathcal{E}^a$ be the respective optimal topology. For the fixed topology $z^*_{e}$, we solve the following original nonlinear, nonconvex program (without integer variables): 
\vspace{-0.4cm}

\begin{subequations} \label{eq:AC_feas}
\allowdisplaybreaks
\begin{align}
\mathcal{O}^*_{feas} := & \min \ \mathcal{O}(f_i^p, l_i^p, l_i^q) \\ \textbf{s.t.} \ & \label{eq:obj_cons}\mathcal{O}(f_i^p, l_i^p, l_i^q) \leq \mathcal{O}^*_{lb}(1+\delta), \\
& \mathrm{Constraints \ \eqref{pbalance}-\eqref{Qloss}}, \\
& z_{e} = z^*_{e} \ \forall e \in \mathcal{E}^a.
\end{align}
\end{subequations}%

where ${\small\mathcal{O}(f_i^p, l_i^p, l_i^q)}$ represents the objective function \eqref{obj}.
Constraint \eqref{eq:obj_cons} specifies that the objective function cost must be within a small percentage $\delta$ of the lower bound $\mathcal{O}^*_{lb}$, where $\delta$ is a specified parameter (3\% for testing). Thus, formulation \eqref{eq:AC_feas} guarantees a feasible solution (close to global optimum) if there exists one for the specified $\delta$. A similar approach has been shown to be effective for OPF-based problems in \cite{molzahn2017laplacian}. The properties of the optimality gap of {\color{black}the test case (single area IEEE RTS-96 system)} from these studies, summarized in Table~\ref{tb:nlgap}, suggest that the relaxed solution is always within 3\% of the optimal solution, indicating that the relaxation is empirically tight to the original MINLP. It is also noteworthy to mention that the convergence time and quality of the local solver (Knitro 10.2.1) solutions were tremendously improved by solving the formulation \eqref{eq:AC_feas}. For instance, we observed gaps up to 70\% (instead of 3\%) by solving formulation \eqref{eq:AC_feas} without the objective-cost constraint in \eqref{eq:obj_cons}.

\begin{table}[!htp]
  \captionsetup{font=footnotesize}
  \centering
  \captionof{table}{Optimality gaps for {
  \color{black}the RTS-96 system} between the lower bound ($\mathcal{O}^*_{lb}$) and the feasible solution ($\mathcal{O}^*_{feas}$) recovered for the original nonconvex model. Values shown are evaluated over various geo-electric field orientations.}
  \begin{tabular}{ccccc}
  \toprule
    &\multicolumn{4}{c}{Optimality Gap (\%)}\\
    \cmidrule(lr){2-5}
    Strength (V/mile)&Avg.&Min.&Max.&Std. Dev.\\
    \midrule
    12&0.9&0.01&2.99&1.4\\
    14&1.8&0.01&3.00&1.5\\
  \bottomrule
  \end{tabular}
  \label{tb:nlgap}
  \vspace{-0.2cm}
\end{table}%

{\color{black} Given the optimality gaps observed here, it is worth noting that the smaller cost fluctuations (up to 3\% in Figure \ref{fig:obj_all}) could be due to the relaxations.
}



\section{Conclusions}
\label{Sec:conc}

We formulated a detailed topology control optimization model to mitigate the impacts of GMD on electrical transmission systems. The mathematical formulation minimizes the total generation dispatch and load shedding subject to nonconvex AC power flow physics, effects of geomagnetically-induced currents on transformer heating and transformer reactive power consumption. Further, we leveraged recently developed convex relaxation approaches to handle the nonlinearites due to AC transmission switching and GIC constraints, which we subsequently observed to provide near global optimum solutions. 

While this paper has made contributions in showing that switching can mitigate the impacts of GMD events, there remain a number of important future directions. For example, new algorithms are needed to solve larger problems. Here, the ACOTS with GMD is naturally posed as a 2-stage program with topology decisions in the master problem. Thus decomposition algorithms, like Benders', are a natural direction to consider. Second, based on our empirical observations, convex relaxation solutions are often tight, thus local search techniques, like meta-heuristics and state-of-the-art global search methods, could yield high quality solutions quickly \cite{nagarajan2016dtmc,nagarajan2017adaptive}. In addition, there are a number of modeling enhancements that need to be considered. For example, integration of  N-1 security (contingency) constraints are important to increase the resiliency of transmission systems under GMD extreme events. Moreover, capturing other effects of GMD on transformers, the modeling of time-extended variations in geo-electric field strengths will be important. Further, there is often uncertainty in predictions of direction and strength of the GMD event, thus it will be important to development methods that produce solutions that are robust to errors in predictions. Finally, this paper modeled GSU and auto transformers, but there are other types of transformers like GWye-GWye Auto and Delta-wye that will need to be modeled.

\section*{Acknowledgment}
The work was funded by the Center for Nonlinear Studies at LANL, the Defense Threat Reduction Agency project \textit{Advancing Knowledge of Networks for Understanding Robustness}, and the LANL laboratory directed research and development project \textit{Impacts of Extreme Space Weather Events on Power Grid Infrastructure: Physics-Based Modeling of Geomagnetically Induced Currents (GICs) During Carrington-Class Geomagnetic Storms}. It was carried out under the auspices of the NNSA of the U.S. DOE at LANL under Contract No. DE-AC52-06NA25396. 

\ifCLASSOPTIONcaptionsoff
  \newpage
\fi

\bibliographystyle{IEEEtran}
\bibliography{references.bib}

\end{document}